\documentclass[a4paper,fleqn]{cas-sc}
\usepackage[utf8]{inputenc}
\usepackage[numbers]{natbib}
\usepackage{microtype}
\usepackage[ruled,vlined]{algorithm2e}
\usepackage[section]{placeins} 
\usepackage{caption}
\usepackage{pifont}
\ExplSyntaxOn
\cs_set:Npn \__first_footerline:
{
  \group_begin:
  \small
  \sffamily
  \ifnum\theblind>0\relax
  \else
    \__short_authors: :~
  \fi
  { \rmfamily \itshape Preprint}
  \group_end:
}
\ExplSyntaxOff

\begin{document}
\shorttitle{Scalable Active Metamaterials for Shape-Morphing}
\shortauthors{J. Cui and W. Chen}
\title[mode=title]{Scalable Active Metamaterials for Shape-Morphing}
\author[1]{Jipeng Cui}
\cormark[1]
\ead{jipengcui@tamu.edu}
\author[1]{Wei~``Wayne'' Chen}
\cormark[1]
\ead{w.chen@tamu.edu}
\affiliation[1]{organization={J. Mike Walker ’66 Department of Mechanical Engineering, Texas A\&M University},
  city={College Station}, postcode={77843}, state={TX}}
\cortext[1]{Corresponding authors}
\begin{abstract}
Shape-morphing materials allow structures such as morphing wings, soft robots, and deployable devices to change their shape on demand without mechanical joints or motors. Aperiodic active metamaterials, in which every unit is designed individually, can realize complex programmed shape changes, but their design by structural optimization requires repeated analyses of the full structure, at a cost that grows rapidly with the number of units. This paper presents a scalable active metamaterial (SAM) framework that designs such structures without simulating the assembled structure. The design domain is discretized into unit cells composed of a stiff frame with compliant hinges and a soft, thermally active infill. The stiff frame limits the interaction between neighboring unit cells, which approximately decouples the design problem into two scales. At the macroscale, the deformation required for each unit cell is obtained by a constrained Laplacian mesh-editing formulation with soft data-driven feasibility constraints. At the microscale, the infill geometry of every unit cell is generated by a conditional diffusion model trained on finite-strain finite element simulations of unit cells. All designs are verified by geometrically nonlinear finite element simulations of the assembled structures. The design cost grows approximately linearly with the number of unit cells. The framework is demonstrated on single-point actuation, multi-point control, and shape matching tasks. We further analyze the sources of the remaining error and discuss possible ways to mitigate it. This work establishes a foundation for the systematic, scalable design of highly complex shape-morphing active metamaterials and for understanding the mechanisms governing their target-matching accuracy.

\end{abstract}
\maketitle
\hypersetup{pdfauthor={Jipeng Cui and Wei Chen},
  pdftitle={Scalable Active Metamaterials for Shape-Morphing},
  pdfsubject={Computer Methods in Applied Mechanics and Engineering},
  pdfkeywords={Active Metamaterial; Scalability; Shape-Morphing; Multi-Scale Design; Programmable Material Systems}}

\section{Introduction}\label{sec:introduction}
\begin{figure}[pos=!htbp]
    \centering
    \includegraphics[width=0.95\linewidth]{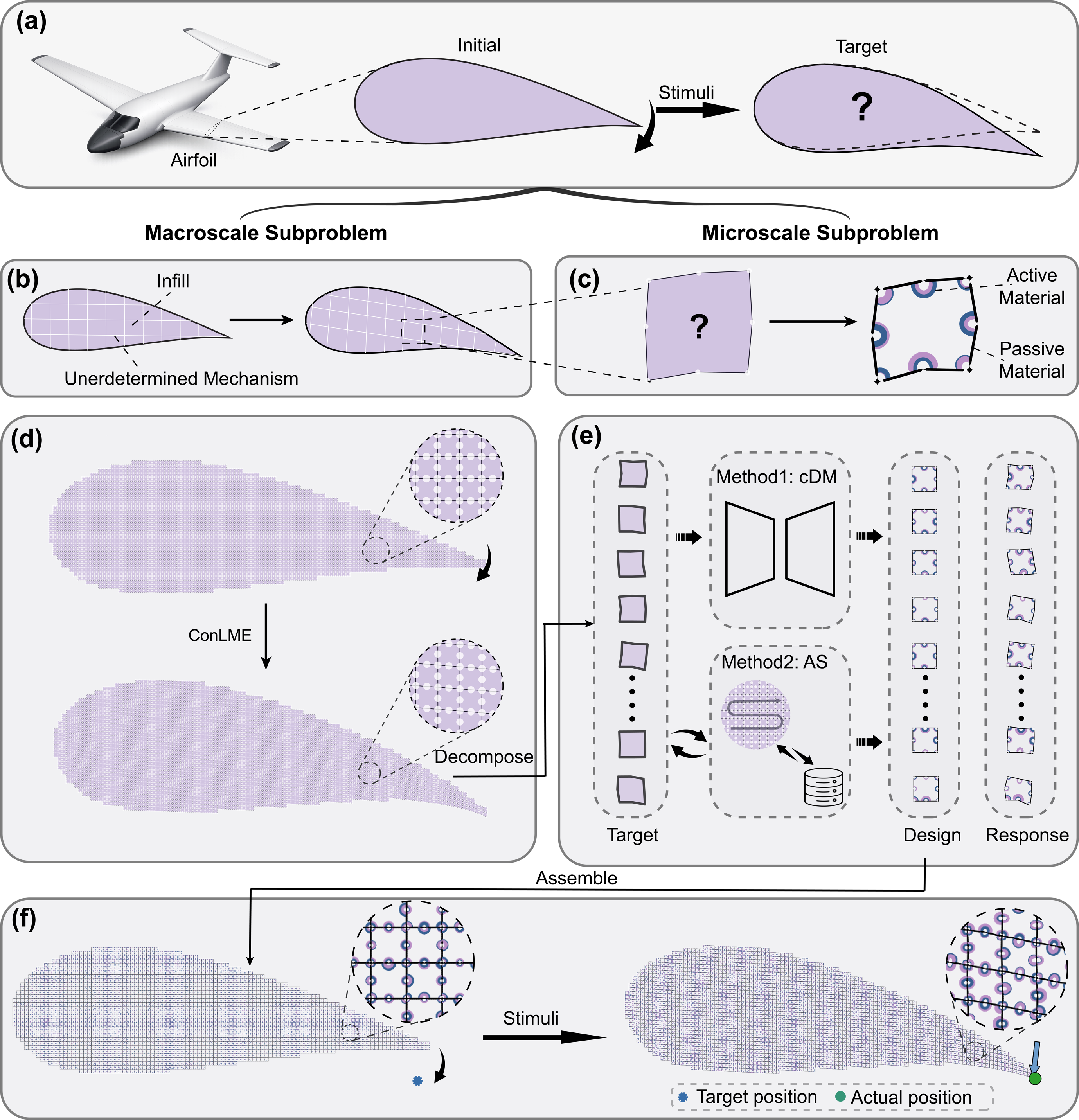}
    \caption{\textbf{ Overview of scalable active metamaterial (SAM) design framework for shape-morphing.}
\textbf{a,} Example shape-morphing task illustrated using an airfoil deformation. Given an initial configuration, the structure is required to deform into a prescribed target shape under external stimulation.
\textbf{b and c,} The design problem is decomposed into two subproblems. The macroscale subproblem in \textbf{b} determines the deformation of free vertices in an underdetermined mechanism given prescribed displacements of selected handle vertices. The microscale subproblem in \textbf{c} seeks the infill geometry of each unit cell that realizes the desired local deformation.
\textbf{d,} Solution to the macroscale subproblem by ConLME. The structure is treated as a mesh, and the global deformation is solved through constrained mesh editing while keeping the unit-cell configurations close to those in the dataset.
\textbf{e,} Solution to the microscale subproblem by cDM or AS. The solved macroscale mesh in \textbf{d} is decomposed into individual unit cell deformation targets, and the inverse design of each unit cell is performed using either a conditional diffusion model or an adjustable search strategy.
\textbf{f,} The designed unit cells shown in \textbf{e} are assembled into the macroscale architecture. Finite element analysis under external stimulation verifies whether the resulting structure achieves the target deformation.}
    \label{fig:overview}
\end{figure}

Active shape-morphing materials \cite{bai_dynamically_2022, meeussen_multistable_2023, siefert_bio-inspired_2019} are engineered materials whose properties are deliberately designed to produce programmable deformation in response to remote stimuli \cite{wang_performance_2024}. Their controllable deformation behavior enables a wide range of applications, including reconfigurable structures \cite{xia_electrochemically_2019}, soft robotics \cite{alapan_reprogrammable_2020}, and minimally invasive medical devices \cite{kim_ferromagnetic_2019}. For example, the wings of an unmanned aerial vehicle (UAV) can be deformed in flight to achieve optimal aerodynamic control (Fig.~\ref{fig:overview}a). However, conventional shape-morphing materials are often limited by restricted design freedom, making it difficult to adapt their deformation behavior to complex, task-specific application scenarios. Metamaterials \cite{wang_mechanical_2022, cui_novel_2023}, whose mechanical properties are governed by microstructural geometry rather than chemical composition, offer a promising route to overcoming this limitation by encoding diverse deformation modes into their architecture. The integration of active stimuli-responsive mechanisms into such architected materials gives rise to active shape-morphing metamaterials (ASMMs) \cite{dudek_shape-morphing_2025}, which enable more flexible and precise programmability of shape change. However, the coupling between the microstructural geometry and material response makes it a fundamental challenge to simultaneously achieve large, programmable deformation while maintaining tractable design complexity in ASMMs.

Addressing this challenge requires a careful examination of how deformation behavior is encoded through structural design in metamaterials. Existing design strategies can be broadly categorized into periodic and aperiodic approaches, each with distinct trade-offs between design complexity and deformation programmability. 

A dominant paradigm is the periodic approach, in which the spatial arrangement of units is treated as a periodic extension \cite{ma_deep_2022} of a single repeating cell, with design efforts focused primarily on optimizing the microstructural geometry within a prescribed unit domain \cite{shaikeea_toughness_2022}, referred to as the \textit{unit cell}. By reconfiguring the internal architecture within this domain, a variety of prescribed deformation behaviors can be achieved, including tailored linear deformation \cite{wenz_designing_2021}, negative Poisson's ratio responses \cite{skarsetz_programmable_2022}, and multistable deformation \cite{greenwood_soft_2025}. Automated design methods, such as topology optimization \cite{zhao_encoding_2023} and data-driven approaches \cite{lee_data-driven_2024}, have further improved the efficiency of generating such periodic metamaterials. The key advantage of this paradigm lies in its significant reduction of design complexity, which greatly improves computational tractability. However, constraining the design to a predefined unit domain and replicating it periodically fundamentally limits the range of achievable deformations, often resulting in structures capable of only simple, globally uniform behaviors such as bending or uniform expansion. 

Removing the constraint of periodicity greatly expands the design space, allowing the geometries and properties of individual units to vary freely across the domain and substantially increasing the dimensionality of the configuration space \cite{zaiser_disordered_2023, bonfanti_automatic_2020}. However, the design complexity afforded by aperiodicity is essentially unbounded, rendering the problem intractable without additional constraints. To manage this complexity, existing approaches can be broadly grouped into three design frameworks: single-unit aperiodic tessellation, hierarchical design, and integrated design.

Single-unit aperiodic tessellation arranges the same building block aperiodically to introduce spatial heterogeneity, thereby enhancing design flexibility while keeping the overall design complexity manageable \cite{jung_aperiodicity_2024, imediegwu_mechanical_2023, liu_growth_2022, coulais_combinatorial_2016}. Nevertheless, the diversity of achievable properties through aperiodic single-unit tiling remains inherently limited. Aperiodic tilings impose strict geometric compatibility requirements on the unit cell, precluding arbitrary cell geometries and thus constraining the accessible design and property space.

To access a broader range of unit configurations, hierarchical design employs scale separation to hierarchically decompose the design problem. The desired mechanical response is prescribed at the macroscale, and a corresponding microstructure is assigned to each unit cell to satisfy the local property requirements \cite{wang_deep_2020, wang_ih-gan_2022}. However, this framework implicitly relies on the assumptions of continuum mechanics, under which material properties such as stiffness are defined as smoothly varying fields. While relaxing these assumptions can still yield viable designs in engineering practice, stronger spatial heterogeneity, characterized by rapid variation in material properties across the domain, necessitates additional treatments, such as those developed for functionally graded materials \cite{lu_making_2014, xu_planar_2016}. Such treatments inevitably reintroduce constraints on the design freedom by requiring material properties to vary smoothly across the domain, which limits the achievable degree of spatial heterogeneity.

A more general approach circumvents this smoothness requirement by designing the entire structure simultaneously, a strategy referred to as integrated design. Representative examples include kirigami and origami-based materials, which reformulate structural design as a geometric problem, reducing design complexity while retaining a high degree of configurational freedom \cite{li_geometric_2024, tang_programmable_2017}. In these systems, the size, distribution, and arrangement of cuts and folds collectively govern the deformation response. However, the underlying design variables, such as cut patterns and fold lines, are often highly coupled and treated as purely geometric primitives, typically determined through iterative optimization \cite{choi_programming_2019}.

Although these design frameworks represent the primary strategies for aperiodic metamaterial design, their extension to ASMM reveals inherent limitations that collectively hinder scalability. The single-unit aperiodic tessellation approach is fundamentally constrained in its ability to achieve complex deformation modes, limiting its applicability to tasks requiring rich spatial programmability. Hierarchical design typically discretizes the domain into bar-based structures admitting only axial deformations \cite{walker_computational_2024}, an assumption that substantially simplifies the design framework and remains valid when local deformations are small. However, active materials are generally soft and undergo large deformations, under which significant non-axial effects arise between adjacent units (see Appendix~\ref{app:discretization} for details), compromising the overall deformation behavior of the assembled structure. This is further supported by the non-axial deformation observed in the experimental results of~\cite{liwei_2025}. Integrated design methods, on the other hand, generally require high-fidelity numerical simulations to obtain the deformed configuration \cite{wang_physics-aware_2023, sim_electromagnetic_nodate}, substantially reducing design efficiency and rendering the approach computationally prohibitive at high resolutions. As a result, all three approaches become increasingly difficult to scale as structural resolution grows, and scalability has long remained a major obstacle to their practical deployment.

To address the scalability challenge, we propose a general scalable active metamaterial (SAM) design framework built on three key contributions. First, we introduce a new design paradigm for aperiodic ASMMs based on a hybrid stiff--soft unit cell discretization, in which the stiff members limit inter-unit interactions while the deformation kinematics of each unit cell implicitly encode the local structural response, circumventing the limitation of the axial-deformation assumption in bar-based methods and eliminating the need for global structural simulation. This enables natural scale separation between the macro- and microscale design subproblems. Second, at the macroscale, we develop a constrained Laplacian mesh editing (ConLME) method that incorporates data-driven constraints, enabling efficient, physically consistent global deformation without expensive physical simulations. Third, at the microscale, we introduce a conditional diffusion model-based inverse design approach, together with a retrieval-based adjustable search strategy that serves as a deterministic reference; the diffusion model is far more efficient at high resolution. The overall framework achieves computational cost that scales approximately linearly with structural resolution---a property that, to our knowledge, has not been demonstrated in existing ASMM design frameworks. Together, these components enable efficient, accurate, and generalizable design of ASMMs, validated on representative deformation tasks including single-point actuation, multi-point control, and curve matching. All designs are verified by geometrically nonlinear finite element simulations of the assembled structures; for physically feasible target deformations, a constraint shared across all existing ASMM design approaches, the mean relative errors of the examples lie between 3\% and 29\%.
\section{Methods}\label{sec:methods}
We first introduce the SAM design framework and then describe the macroscale and microscale design methods. The unit cell design, the finite element modeling, the data generation, the evaluation metrics, and the computational settings are detailed in Appendices~\ref{sec:unit-cell}, \ref{app:fem}, \ref{sec:data-generation}, \ref{sec:metrics}, and~\ref{app:computational-setup}, respectively.

\subsection{Scalable active metamaterial (SAM) design framework for shape-morphing}\label{sec:framework}

The goal of the SAM framework is to transform a given initial shape into a target configuration using active metamaterials (Fig.~\ref{fig:overview}a). We discretize the entire design domain into unit cells with a hybrid architecture composed of soft and stiff materials. Since the deformation of the stiff material is negligible relative to that of the soft material, we treat the stiff members as rigid and neglect their contribution to the overall deformation in the analysis. This material contrast enables a natural decomposition of the problem into two scales: macroscale and microscale. At the macroscale, rigid bars partition the domain into equal-sized square unit cells (Fig.~\ref{fig:overview}b), where adjacent cells are connected by hinged rigid bars, forming an underdetermined linkage mechanism. This structure admits a mesh representation $\boldsymbol{M}=(\mathcal{V},\mathcal{E})$, where $\mathcal{V}$ denotes all vertices, each corresponding to a hinge joint in the linkage, and $\mathcal{E}$ denotes the edges representing the connecting rigid bars. The coordinates of all vertices are vectorized as $\boldsymbol{v} \in \mathbb{R}^{2N_v}$ when used in linear algebraic formulations, where $N_v = |\mathcal{V}|$ is the total number of vertices. At the microscale (Fig.~\ref{fig:overview}c), each unit cell contains a soft material infill with a specific geometrical structure $\boldsymbol{x}$ that deforms in response to external stimuli, thereby driving the relative motion of the rigid bars and enabling global shape morphing. In this work, the infill material exhibits thermally induced mechanical deformation. The detailed parameterization is provided in Appendix~\ref{sec:unit-cell}. Under this setup, the micro- and macro-scales decouple naturally. By further assuming that adjacent unit cells exhibit identical displacements at their shared interfaces in response to external stimuli, we decompose the overall design problem into two independent subproblems, each solvable without explicitly modeling the interactions between unit cells. This decoupling is an approximation. The interaction that remains between the unit cells of an assembled structure is discussed in Section~\ref{sec:error-sources} and quantified in Appendix~\ref{app:assembly}.

To solve the macroscale subproblem, we develop Constrained Laplacian Mesh Editing (ConLME) (Fig.~\ref{fig:overview}d). Based on the assumption that deformation of the rigid components is negligible, the macroscale subproblem reduces to a mesh deformation problem that requires no physical simulation. Given a set of handle vertices whose positions after deformation are prescribed according to the target shape, the macroscale subproblem seeks the positions of the remaining free vertices. Specifically, given the prescribed positions of handle vertices $\boldsymbol{v}^t_\mathrm{handle}$, we formulate this as the following optimization problem:
\begin{equation}
    \boldsymbol{v}^*_\mathrm{free} = \arg\min_{\boldsymbol{v}_\mathrm{free}}\|\boldsymbol{v}^t_{\mathrm{handle}}-\boldsymbol{v}_{\mathrm{handle}}(\boldsymbol{v}_\mathrm{free})\|_2,
    \label{Eq:initial optimization problem}
\end{equation}
where $\boldsymbol{v}_\mathrm{free}$ denotes the coordinates of the free vertices and $\boldsymbol{v}_\mathrm{handle}$ denotes the deformed positions of the handle vertices. The full vertex coordinates $\boldsymbol{v}^*$, assembled from the optimized free vertices $\boldsymbol{v}^*_\mathrm{free}$ and the handle vertices $\boldsymbol{v}_\mathrm{handle}$, are then decomposed into individual unit cells. Each cell is represented by a specific equilateral octagonal configuration (Fig.~\ref{fig:overview}c), which can also be fully characterized by its eight internal angles $\boldsymbol{\theta}$. The solution to this optimization problem must satisfy three criteria. First, the mesh topology must be preserved to ensure the stability of local relative coordinates. Second, the lengths of mesh edges should remain unchanged. Third, the final configuration of each unit cell should lie within the feasible region reachable by the local design method. To enforce these criteria, we modify the Laplacian surface editing \cite{sorkine_laplacian_2004, igarashi_as-rigid-as-possible_2005} approach by introducing soft constraints into the original optimization problem, yielding the ConLME formulation. Consequently, the original problem in Eq.~\eqref{Eq:initial optimization problem} is transformed into the minimization of a weighted energy function:
\begin{equation}
    \begin{aligned}
        \boldsymbol{v}^*_\mathrm{free}= \arg \min_{\boldsymbol{v_\mathrm{free}}} w_LE_L(\boldsymbol{v})+w_SE_S(\boldsymbol{v})+w_tE_t(\boldsymbol{v}_{\mathrm{handle}}(\boldsymbol{v}_\mathrm{free});\boldsymbol{v}^t_{\mathrm{handle}}).
    \end{aligned}
    \label{Eq:Optimization problem}
\end{equation}
The three terms serve distinct roles: the \textit{Laplacian energy} $E_L(\boldsymbol{v})$ promotes smooth deformation by keeping vertices close to the centroid of their neighbors; the \textit{configuration consistency energy} $E_S(\boldsymbol{v})$ keeps each unit cell close to its nearest configuration in the preconstructed dataset $\mathcal{D}$, which encourages both edge length invariance and the feasible design range of the microscale subproblem; and the \textit{target matching energy} $E_t(\boldsymbol{v})$ drives the handle vertices toward their prescribed target positions. With weights $w_L$, $w_S$, and $w_t$ (selected as described in Appendix~\ref{sec:weight-selection}), the second and third terms act as soft constraints: they penalize, but do not exclude, stretched edges and configurations far from the dataset, so the solution is checked afterwards with the dissimilarity and edge length deviation diagnostics of Section~\ref{sec:error-sources}. For fixed reference configurations, the resulting objective is a positive definite quadratic whose minimizer is obtained by solving a linear system; alternating this solve with the retrieval of reference configurations monotonically decreases the objective, which yields an efficient solution to the macroscale subproblem. Definitions of the three terms and the derivation of the linear system are given in Section~\ref{sec:conlme}.

The microscale subproblem performs inverse design of the internal infill geometry $\boldsymbol{x}$ for each unit cell, given the target configuration angles $\boldsymbol{\theta}$ obtained from the macroscale solution. This problem is essentially equivalent to the inverse design of conventional passive metamaterial unit cells \cite{wang_ih-gan_2022, kollmann_deep_2020, mao_designing_2020} and can be addressed within an acceptable error tolerance using standard generative methods (Fig.~\ref{fig:overview}e). The designed infills are then assembled to form the final structure that performs the target deformation (Fig.~\ref{fig:overview}f). We address this inverse design task using two alternative approaches, each offering distinct trade-offs between computational cost and design flexibility.

The first is an adjustable search (AS) strategy, in which candidate unit cell configurations are retrieved from a preconstructed dataset $\mathcal{D}$ and iteratively substituted into the global structure, followed by re-solving the macroscale subproblem. This process repeats until all unit cell configurations are determined. Details of the dataset $\mathcal{D}$ and the full procedure are provided in Appendix~\ref{sec:data-generation} and Section~\ref{sec:as}, with geometric parameterization in Appendix~\ref{sec:unit-cell}.

The second approach employs a conditional diffusion model (cDM) \cite{bastek_inverse_2023, ho_imagen_2022, ho_denoising_2020} to learn the conditional distribution of infill geometry $\boldsymbol{x}$, denoted $p(\boldsymbol{x}|\boldsymbol{\theta})$. During inference, candidate infill designs are generated through an iterative denoising process conditioned on the prescribed unit cell configuration $\boldsymbol{\theta}$. The model architecture and implementation details are provided in Section~\ref{sec:cdm}.

\subsection{Constrained Laplacian mesh editing}
\label{sec:conlme}
This section defines the three energy terms in Eq.~\eqref{Eq:Optimization problem}, and derives the linear system that minimizes their weighted sum; weight selection and iterative convergence are analyzed in Appendix~\ref{sec:weight-selection}.

The Laplacian energy is defined as
\begin{equation}
    \begin{aligned}
        E_L(\boldsymbol{v}) = \sum_{i\in \mathcal{V}_L}\left\|\boldsymbol{v}_i-\frac{1}{n_i}\sum_{j\in \mathcal{N}_i}\boldsymbol{v}_j\right\|^2,
    \end{aligned}
    \label{Eq:EL}
\end{equation}
where $\boldsymbol{v}$ is the vectorization of all vertices $\mathcal{V}$, $\boldsymbol{v}_i\in \mathcal{V} $ denotes the coordinate of vertex $i$, $n_i$ is the number of its neighbors, $\mathcal{N}_i$ is the corresponding neighbor set, and $N_v=\left |\mathcal{V}\right|$. The sum runs over $\mathcal{V}_L\subseteq\mathcal{V}$, which excludes the isolated hinge vertices at the corners of boundary cells to prevent the boundary from collapsing. This term penalizes the deviation of each remaining vertex from the centroid of its neighbors, thereby preserving mesh topology and promoting smooth deformation. In matrix form, this energy can be written compactly as
\begin{equation}
    E_L(\boldsymbol{\boldsymbol{v}}) = \|\mathbf{L}_L\boldsymbol{v}\|^2,
    \label{term1}
\end{equation}
where $\mathbf{L}_L$ is the normalized graph Laplacian matrix with the rows of the excluded vertices set to zero. Since $\boldsymbol{v}\in\mathbb{R}^{2N_v}$ stacks both coordinates of every vertex, all vertex-level matrices in this section act on the two coordinates independently, i.e., they are Kronecker products with $\mathbf{I}_2$; the factor $\otimes\,\mathbf{I}_2$ is omitted for brevity.

The configuration consistency energy is defined as
\begin{equation}
    E_S\big(\boldsymbol{v}; \{c_i, \mathbf{R}_i, \boldsymbol{t}_i\}\big) = \sum_{i=1}^{N_c}\sum_{j=1}^{8}\left\|\boldsymbol{v}_j^{(i)} - \left(\mathbf{R}_i\hat{\boldsymbol{v}}_j^{(c_i)} + \boldsymbol{t}_i\right)\right\|^2,
    \label{Eq:ES}
\end{equation}
where $\boldsymbol{v}_j^{(i)}$ denotes the $j$-th vertex of cell $i$, $N_c$ is the total number of unit cells, and $\hat{\boldsymbol{v}}_j^{(c_i)}$ is the $j$-th vertex of the configuration $c_i \in \mathcal{D}$ assigned to cell $i$, placed at that cell by the rigid transformation $(\mathbf{R}_i, \boldsymbol{t}_i)$. For a given deformed mesh, each cell is assigned the configuration that fits it best, i.e., Eq.~\eqref{Eq:ES} is minimized over the assignment and the alignment, which decouples into one problem per cell:
\begin{equation}
    (c_i^*, \mathbf{R}_i^*, \boldsymbol{t}_i^*) = \arg\min_{\substack{1 \leq c \leq |\mathcal{D}|,\ \mathbf{R}\in \mathrm{SO}(2),\ \boldsymbol{t}\in\mathbb{R}^2}} \sum_{j=1}^{8}\left\|\boldsymbol{v}_j^{(i)} - \left(\mathbf{R}\,\hat{\boldsymbol{v}}_j^{(c)} + \boldsymbol{t}\right)\right\|^2.
    \label{Eq:retrieval}
\end{equation}
We write $\hat{\boldsymbol{v}}_j^{(i)} = \mathbf{R}_i^*\hat{\boldsymbol{v}}_j^{(c_i^*)} + \boldsymbol{t}_i^*$ for the resulting aligned reference position and abbreviate $E_S(\boldsymbol{v}) = E_S\big(\boldsymbol{v}; \{c_i^*, \mathbf{R}_i^*, \boldsymbol{t}_i^*\}\big)$. For a given candidate $c$, the inner minimization over $(\mathbf{R}, \boldsymbol{t})$ is the Procrustes problem of optimally superimposing two ordered point sets, whose closed-form solution is classical \cite{schonemann_generalized_1966, umeyama_least-squares_1991}: the optimal translation matches the centroids of the two configurations, and the optimal rotation follows from the cross-covariance matrix of the centroid-subtracted vertices. The minimization over $c$ is then an exhaustive search over $\mathcal{D}$, which is evaluated in closed form for all candidates at once. Reflections are excluded ($\mathbf{R}\in\mathrm{SO}(2)$), since a mirrored configuration corresponds to a different infill design.

This term encourages each unit cell to remain close to a configuration in $\mathcal{D}$, avoiding out-of-distribution solutions and implicitly constraining mesh edge lengths to remain near their initial values. To write this energy in matrix form for a fixed set of aligned references, we regroup the double sum in Eq.~\eqref{Eq:ES} by global vertex. Let $\mathcal{U}_k$ denote the set of unit cells containing vertex $k$, $d_k = |\mathcal{U}_k|$ the number of such cells, and, with a slight abuse of notation, $\hat{\boldsymbol{v}}_k^{(i)}$ the reference position that the aligned configuration of cell $i\in\mathcal{U}_k$ assigns to vertex $k$. Then
\begin{equation}
    E_S(\boldsymbol{v}) = \sum_{k=1}^{N_v}\sum_{i\in\mathcal{U}_k}\|\boldsymbol{v}_k - \hat{\boldsymbol{v}}_k^{(i)}\|^2 = \boldsymbol{v}^T\mathbf{L}_S\boldsymbol{v} - 2\,\boldsymbol{v}^T\boldsymbol{b}^{\mathcal{D}} + c_S,
    \label{term2}
\end{equation}
where $\mathbf{L}_S = \mathrm{diag}(d_1, d_2, \ldots, d_{N_v})$, $c_S = \sum_{k}\sum_{i\in\mathcal{U}_k}\|\hat{\boldsymbol{v}}_k^{(i)}\|^2$ is independent of $\boldsymbol{v}$, and $\boldsymbol{b}^{\mathcal{D}}$ is a vector whose $k$-th entry is the sum of the reference positions of vertex $k$ over all cells containing it:
\begin{equation}
    \boldsymbol{b}^{\mathcal{D}}_k = \sum_{i\in\mathcal{U}_k} \hat{\boldsymbol{v}}_k^{(i)}.
\end{equation}
Equivalently, $E_S(\boldsymbol{v}) = \sum_k d_k\|\boldsymbol{v}_k - \boldsymbol{b}^{\mathcal{D}}_k/d_k\|^2 + \mathrm{const}$, i.e., each vertex is attracted to the average of its reference positions with a weight equal to the number of cells sharing it.

The target matching energy is defined as
\begin{equation}
    \begin{aligned}
        E_t(\boldsymbol{v}) = \sum_{\boldsymbol{v}_i\in \mathcal{V}_\mathrm{handle}}\|\boldsymbol{v}_i-\boldsymbol{v}^t_i\|^2,
    \end{aligned}
\end{equation}
where $\mathcal{V}_\mathrm{handle}$ is the set of handle vertices and $\boldsymbol{v}^t_i$ is the target position of handle vertex $i$. Fixed vertices, i.e., those anchored by the boundary conditions, are treated as handle vertices with zero prescribed displacement ($\boldsymbol{v}^t_i$ equal to their initial position) and are therefore included in $\mathcal{V}_\mathrm{handle}$. This term enforces the boundary conditions by driving the vertices with prescribed target positions toward their desired locations. In the matrix form, it could be written as,
\begin{equation}
        \begin{aligned}
        E_t(\boldsymbol{v}) =\|\mathbf{L}_t\boldsymbol{v}-\boldsymbol{b}^t\|^2,
        \label{term3}
    \end{aligned}
\end{equation}
where $\mathbf{L}_t = \mathrm{diag}(\delta_1, \delta_2, \ldots, \delta_{N_v})$ is a diagonal selection matrix with
\begin{equation}
    \delta_j = \begin{cases} 1 & \text{if } \boldsymbol{v}_j \in \mathcal{V}_\mathrm{handle}, \\ 0 & \text{otherwise,} \end{cases}
\end{equation}
and $\boldsymbol{b}^t$ is the target position vector whose $j$-th entry equals the prescribed target position $\boldsymbol{v}_j^t$ if $\boldsymbol{v}_j \in \mathcal{V}_\mathrm{handle}$ and zero otherwise. This term penalizes the deviation of handle vertices from their prescribed target positions.

Substituting Eqs.~\eqref{term1}, \eqref{term2}, and \eqref{term3} into Eq.~\eqref{Eq:Optimization problem}, it can be written in algebraic form as
\begin{equation}
    \begin{aligned}
        E(\boldsymbol{v}) = w_L\|\textbf{L}_L\boldsymbol{v}\|^2+w_S\left(\boldsymbol{v}^T\textbf{L}_S\boldsymbol{v}-2\,\boldsymbol{v}^T\boldsymbol{b}^{\mathcal{D}}+c_S\right)+w_t\|\textbf{L}_t\boldsymbol{v}-\boldsymbol{b}^t\|^2.
    \end{aligned}
    \label{fullform-energy}
\end{equation}

For fixed reference configurations $\{\hat{\boldsymbol{v}}^{(i)}\}$, each term is quadratic in $\boldsymbol{v}$ and all weighting coefficients are positive, so the optimization problem admits an analytical solution. By differentiating Eq.~\eqref{fullform-energy} with respect to $\boldsymbol{v}$, we obtain
\begin{equation}
    \begin{aligned}
        \frac{\partial E}{\partial \boldsymbol{v}}=2\left[(w_L\textbf{L}_L^T\textbf{L}_L+w_S\textbf{L}_S+w_t\textbf{L}_t^T\textbf{L}_t)\boldsymbol{v}-(w_S\boldsymbol{b}^{\mathcal{D}}+w_t\textbf{L}_t^T\boldsymbol{b}^t)\right].
    \end{aligned}
    \label{full-differential}
\end{equation}

By setting Eq.~\eqref{full-differential} to zero, a concise linear system is obtained:
\begin{equation}
    \begin{aligned}
        \textbf{L}^a\boldsymbol{v}=\boldsymbol{b}^a
    \end{aligned}
    \label{Linear system}
\end{equation}
where
\begin{equation}
    \textbf{L}^a=w_L\textbf{L}_L^T\textbf{L}_L+w_S\textbf{L}_S+w_t\textbf{L}_t^T\textbf{L}_t,
\end{equation}
and
\begin{equation}
    \boldsymbol{b}^a=w_S\boldsymbol{b}^{\mathcal{D}}+w_t\textbf{L}_t^T\boldsymbol{b}^t.
\end{equation}
Because every vertex belongs to at least one unit cell ($d_k \geq 1$), $\textbf{L}_S$ is positive definite, and hence so is $\textbf{L}^a$; the linear system therefore has a unique solution, which is the minimizer of Eq.~\eqref{fullform-energy} for the current reference configurations. Pure Laplacian surface editing \cite{sorkine_laplacian_2004} is used to obtain the initial design.

In general, the reference configurations retrieved by Eq.~\eqref{Eq:retrieval} change after each solve of Eq.~\eqref{Linear system}, so the macroscale subproblem is solved iteratively by alternating two steps: (i) \textit{retrieval}, in which $\boldsymbol{v}$ is fixed and Eq.~\eqref{Eq:retrieval} is solved for every cell; and (ii) \textit{update}, in which the aligned references are fixed and Eq.~\eqref{Linear system} is solved for $\boldsymbol{v}$. Both steps minimize the same energy $w_L E_L(\boldsymbol{v}) + w_S E_S\big(\boldsymbol{v}; \{c_i, \mathbf{R}_i, \boldsymbol{t}_i\}\big) + w_t E_t(\boldsymbol{v})$, now regarded as a function of the vertex coordinates \textit{and} the assignment and alignment variables: step (i) minimizes it exactly over $\{c_i, \mathbf{R}_i, \boldsymbol{t}_i\}$, since $E_L$ and $E_t$ do not depend on these variables and $E_S$ decouples across cells, and step (ii) minimizes it exactly over $\boldsymbol{v}$. Consequently, this energy is monotonically non-increasing over the iterations and, being bounded below by zero, converges. The iteration is terminated when the retrieved configurations no longer change or a prescribed maximum number of iterations is reached.

\subsection{Adjustable search}
\label{sec:as}
The AS strategy performs microscale inverse design by sequentially retrieving unit cell configurations from the dataset $\mathcal{D}$ (Algorithm~\ref{alg:as}). Given a mesh $(\mathcal{V}, \mathcal{E})$, a unit cell dataset $\mathcal{D}$, and an initial fixed cell set $\mathcal{F}$---where a fixed cell is defined as any unit cell containing at least one handle vertex---the macroscale deformation is first resolved via ConLME to obtain an initial vertex configuration $\mathcal{V}_0$. Candidate cells are then traversed sequentially in a prescribed search order; for each cell, the nearest-neighbor configuration is retrieved from $\mathcal{D}$ and aligned to the current cell location by rotating about the axis defined by the known vertex pair, ensuring consistency with the already-fixed vertices. The retrieved configuration is then substituted into the global structure, its vertices become fixed vertices, and ConLME is re-solved with all fixed vertices treated as handle vertices to update the positions of the remaining free vertices. This process repeats until all unit cells have been assigned a configuration.

The search order is determined according to the following principles:
\begin{enumerate}
    \item Unit cells in the fixed cell set $\mathcal{F}$ are prioritized;
    \item Among cells with equal fixed status, those with a greater number of fixed vertices are prioritized;
    \item Among cells satisfying the above conditions equally, those undergoing larger deformations in $\mathcal{V}_0$, measured by the magnitude of the area variation $|\Delta A|$ (Appendix~\ref{sec:weight-selection}), are prioritized.
\end{enumerate}
Specifically, the search begins from the fixed cells at the boundary and proceeds outward. The key advantage of this ordering strategy is that it prioritizes cells whose vertex positions are already well-constrained, ensuring that the macroscale solution is updated with reliable reference points and thereby reducing accumulated error in the remaining free vertices.

\RestyleAlgo{ruled}
\begin{algorithm}[htbp]
\caption{Adjustable Search (AS)}\label{alg:as}
\KwIn{Unit cell vertices $\mathcal{V} = \bigcup_{k=1}^{N_c} \{\boldsymbol{v}^{(k)}_j\}_{j=1}^{8}$ where $\{\boldsymbol{v}^{(k)}_j\}_{j=1}^{8}$ denotes the configuration of the $k$-th unit cell, unit cell dataset $\mathcal{D} = \{\mathcal{C}^{(k)}\}_{k=1}^{|\mathcal{D}|}$ where $\mathcal{C}^{(k)} = \{\hat{\boldsymbol{v}}^{(k)}_j\}_{j=1}^{8}$ is one configuration in the dataset, initial set of fixed cell indices $\mathcal{F}$}
\KwOut{Optimized unit cell vertices $\{\boldsymbol{v}^{(k)*}_j\}_{j=1}^{8}$ for $k = 1,\ldots,N_c$}
Solve Eq.~\eqref{Linear system} to obtain initial vertex positions $\mathcal{V}_0$\;
Determine search order $\mathcal{S} = [s_1, s_2, \dots, s_{N_c}]$ according to the prioritization rules\;
\For{$i = 1$ \KwTo $N_c$}{
    Retrieve nearest-neighbor configuration from dataset:
    $c^* \leftarrow \arg\min_{1 \leq c \leq |\mathcal{D}|} \sum_{j=1}^{8}\|\boldsymbol{v}^{(s_i)}_j - \hat{\boldsymbol{v}}^{(c)}_j\|^2$\;
    Align $\mathcal{C}^{(c^*)}$ to the current cell location via rotation about the known vertex axis\;
    Substitute aligned configuration: $\boldsymbol{v}^{(s_i)}_j \leftarrow \hat{\boldsymbol{v}}^{(c^*)}_j$ for $j = 1,\ldots,8$\;
    Update fixed cell set: $\mathcal{F} \leftarrow \mathcal{F} \cup \{s_i\}$\;
    Re-solve Eq.~\eqref{Linear system} to obtain updated $\mathcal{V}$\;
}
\Return $\boldsymbol{v}^{(k)*}_j \leftarrow \boldsymbol{v}^{(k)}_j$ for $k = 1,\ldots,N_c$, $j = 1,\ldots,8$\;
\end{algorithm}

\begin{figure}[pos=!htbp]
    \centering
    \includegraphics[width=1.0\linewidth]{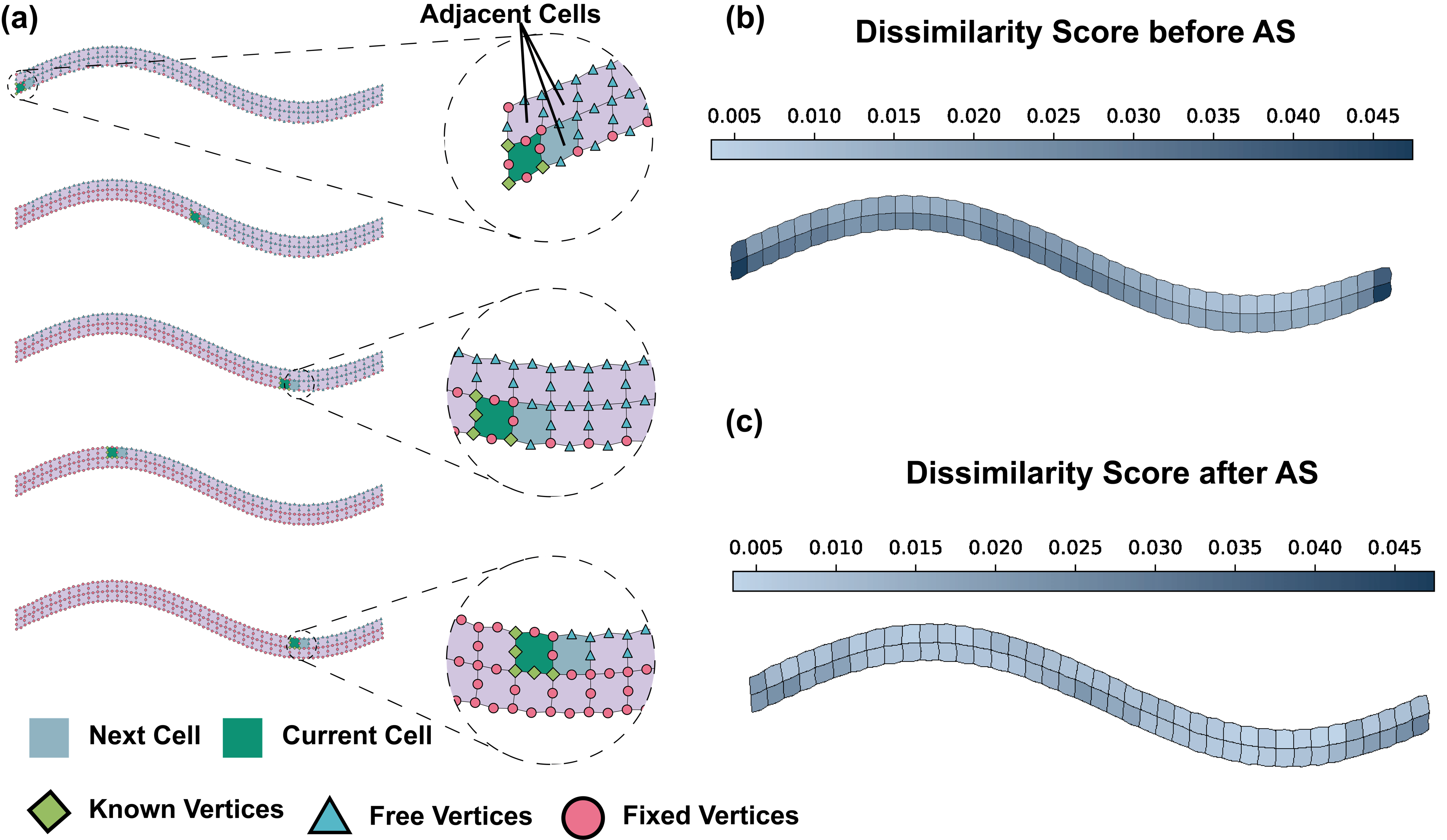}
    \caption{
    \textbf{Adjustable search (AS) method.}
    \textbf{a,} Illustration of the AS procedure.
    \textbf{b and c,} Dissimilarity scores before (\textbf{b}) and after (\textbf{c}) AS for the sinusoidal beam benchmark.
    }
    \label{fig:as}
\end{figure}

Fig.~\ref{fig:as}a illustrates the execution of the AS algorithm step by step. A comparison of dissimilarity scores before and after the AS update (Figs.~\ref{fig:as}b--c) shows a marked reduction in overall dissimilarity, since every unit cell is replaced by a configuration retrieved from the dataset.

\subsection{Conditional diffusion model}
\label{sec:cdm}
Unlike AS, which retrieves existing designs from $\mathcal{D}$, the cDM generates infill geometries by sampling from the learned conditional distribution $p(\boldsymbol{x}|\boldsymbol{\theta})$. A U-Net \cite{ronneberger_u-net_2015} predicts the noise added at each diffusion timestep; following the conditional DDPM settings in \cite{bastek_inverse_2023}, the target configuration is injected throughout the network.

\subsubsection{Fundamentals of the denoising diffusion probabilistic model (DDPM)}

The denoising diffusion probabilistic model (DDPM)~\cite{ho_denoising_2020} defines a forward process that progressively corrupts clean design parameters $\boldsymbol{x}_0$ into pure Gaussian noise $\boldsymbol{x}_T$, and a reverse process in which a deep neural network learns to reconstruct the original design from noise. In the forward process, Gaussian noise $\boldsymbol{\epsilon} \sim \mathcal{N}(\boldsymbol{0}, \boldsymbol{I})$ is added to $\boldsymbol{x}_0$ according to a predefined noise schedule $\{\beta_t \in (0,1)\}_{t=1}^T$, forming a Markov chain:
\begin{equation}
    q(\boldsymbol{x}_{1:T}|\boldsymbol{x}_0) = \prod_{t=1}^T q(\boldsymbol{x}_t|\boldsymbol{x}_{t-1}), \quad
    q(\boldsymbol{x}_t|\boldsymbol{x}_{t-1}) = \mathcal{N}\!\left(\boldsymbol{x}_t;\, \sqrt{\alpha_t}\,\boldsymbol{x}_{t-1},\, (1-\alpha_t)\boldsymbol{I}\right),
\end{equation}
where $\bar{\alpha}_t = \prod_{i=1}^t (1-\beta_i)$, which allows sampling at arbitrary timesteps via:
\begin{equation}
    \boldsymbol{x}_t = \sqrt{\bar{\alpha}_t}\,\boldsymbol{x}_0 + \sqrt{1-\bar{\alpha}_t}\,\boldsymbol{\epsilon}.
\end{equation}
The reverse process begins by sampling $\boldsymbol{x}_T \sim \mathcal{N}(\boldsymbol{0}, \boldsymbol{I})$ and iteratively samples from $p(\boldsymbol{x}_{t-1}|\boldsymbol{x}_t)$ until $\boldsymbol{x}_0$ is reconstructed. This distribution is parameterized by a neural network:
\begin{equation}
    p_{\boldsymbol{\phi}}(\boldsymbol{x}_{t-1}|\boldsymbol{x}_t) = \mathcal{N}\!\left(\boldsymbol{x}_{t-1};\, \boldsymbol{\mu}_{\boldsymbol{\phi}}(\boldsymbol{x}_t, t),\, \sigma_t^2 \boldsymbol{I}\right),
\end{equation}
where $\boldsymbol{\phi}$ denotes the network parameters and the variance is fixed as:
\begin{equation}
    \sigma_t^2 = \frac{1 - \bar{\alpha}_{t-1}}{1 - \bar{\alpha}_t}\,\beta_t.
\end{equation}
Following~\cite{ho_denoising_2020}, the network is trained by directly predicting the added noise rather than maximizing the evidence lower bound (ELBO):
\begin{equation}
    \mathcal{L}(\boldsymbol{\phi}) = \mathbb{E}_{\boldsymbol{\epsilon} \sim \mathcal{N}(\boldsymbol{0},\boldsymbol{I}),\, \boldsymbol{x}_0,\, t \sim \mathrm{Unif}(1,T)}\!\left[\left\|\boldsymbol{\epsilon} - \boldsymbol{\epsilon}_{\boldsymbol{\phi}}(\boldsymbol{x}_t, t)\right\|^2\right].
\end{equation}

\subsubsection{Conditional encoding}

Since the task requires generating unit cell designs conditioned on the target deformed configuration $\boldsymbol{\theta}$, a conditional diffusion model is employed:
\begin{equation}
    q_{\boldsymbol{\phi}}(\boldsymbol{x}_{t-1}|\boldsymbol{x}_t, \boldsymbol{\theta}) = \mathcal{N}\!\left(\boldsymbol{x}_{t-1};\, \boldsymbol{\mu}_{\boldsymbol{\phi}}(\boldsymbol{x}_t, t, \boldsymbol{\theta}),\, \sigma_t^2\boldsymbol{I}\right).
\end{equation}
Following~\cite{rombach_high-resolution_2022}, the condition $\boldsymbol{\theta}$ is injected via cross-attention at multiple scales of the U-Net. No separate condition encoder is used: the min--max normalized angle vector $\boldsymbol{\theta}\in\mathbb{R}^{1\times 8}$ is passed directly as the context of every cross-attention layer, which projects it to keys and values through its own linear maps. At each downsampling block, the feature map is fused with a cross-attention output computed from $\boldsymbol{\theta}$, then processed by residual and attention layers before downsampling. The same conditioning is applied at the bottleneck and at each upsampling block after skip concatenation, ensuring that condition information is continuously re-incorporated throughout both the compression and reconstruction stages.

\subsubsection{Model architecture and training}

The noise-prediction network is a one-dimensional U-Net with four encoder stages, a bottleneck, and a mirrored decoder with skip connections. The diffusion timestep modulates every residual block, and the target configuration $\boldsymbol{\theta}$ is injected by cross-attention at every stage, as described above. The diffusion process uses $T = 1000$ timesteps, and designs are generated by sampling over all $T$ steps. The model is trained on 80\% of the dataset, with the remaining 20\% held out as the test set, and training completes in less than one hour on a single GPU. On the test set, the model reaches a coefficient of determination of 0.91, corresponding to an RMS angle error of $3.41^\circ$. The full architecture, the training settings, and the evaluation procedure are given in Appendix~\ref{app:cdm-model}.

\section{Results}\label{sec:results}
This section evaluates four key characteristics of the proposed SAM framework: scalability, accuracy, generality, and compactness.

\subsection{Linear scalability in cost}\label{sec:scalability}

\begin{figure}[pos=!htbp]
    \centering
    \includegraphics[width=0.8\linewidth]{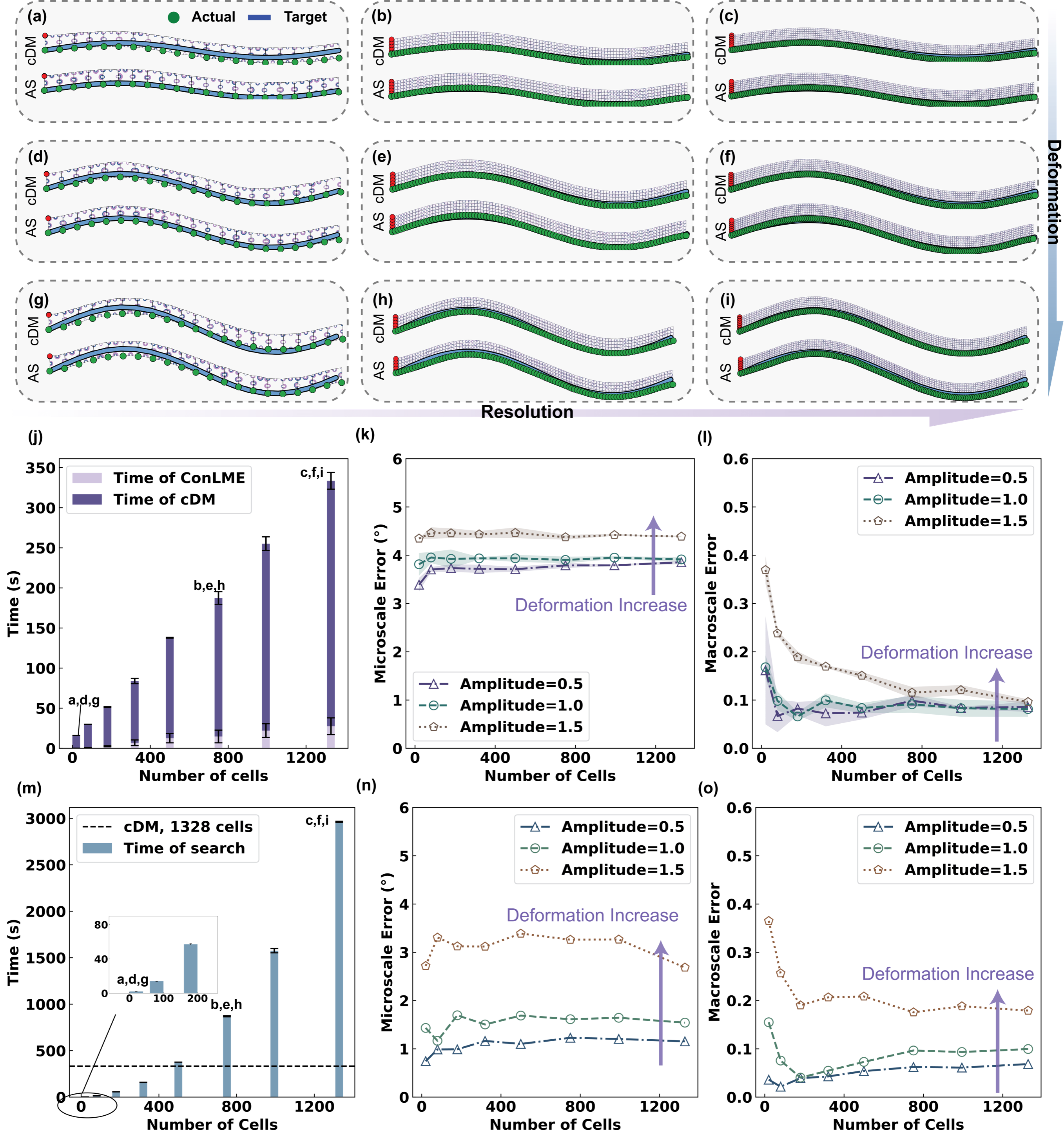}
    \caption{\textbf{Computational efficiency and accuracy of the SAM framework evaluated on the sinusoidal beam benchmark.}
    \textbf{a--i,} Deformation results under different resolutions and deformation magnitudes. From left to right, the resolution increases, and from top to bottom, the deformation magnitude increases. In each panel, the upper plot corresponds to local design using the conditional diffusion model (cDM), while the lower plot corresponds to the adjustable search (AS) strategy. The blue curve denotes the target positions of the handle vertices, green dots represent their deformed positions obtained from finite element simulations, and red dots mark the fixed vertices.
    \textbf{j,} Wall-clock time of the cDM-based SAM under different resolutions, decomposed into the ConLME stage (for the macroscale subproblem) and the cDM stage (for the microscale subproblem). In \textbf{j}--\textbf{l}, each configuration is designed three times with independent diffusion samples; the bars and curves show the mean, and the error bars and shaded bands show one standard deviation. The labels above the bars in \textbf{j} and \textbf{m} indicate the resolutions shown in the deformation panels: 20 (\textbf{a}, \textbf{d}, \textbf{g}), 750 (\textbf{b}, \textbf{e}, \textbf{h}), and 1328 (\textbf{c}, \textbf{f}, \textbf{i}) unit cells.
    \textbf{k,} Microscale performance of the cDM-based SAM measured by the RMS error $\mathrm{RMS}_{\mathrm{micro}}$ between the target angle changes of the unit cells and those simulated for the inverse-designed unit cells under varying resolutions and deformation magnitudes.
    \textbf{l,} Macroscale performance of the cDM-based SAM, measured as the MAE between target positions and the actual deformed positions of handle vertices based on the FEA simulation of the assembled metamaterial structure, divided by the beam width.
    \textbf{m,} Wall-clock time of the AS-based SAM under different resolutions. AS is deterministic, so each configuration is designed once. The dashed line marks the time of the cDM-based SAM with 1328 unit cells, and the inset enlarges the three coarsest designs.
    \textbf{n,} Microscale performance of the AS-based SAM.
    \textbf{o,} Macroscale performance of the AS-based SAM, normalized as in \textbf{l}.}
    \label{fig:scalability}
\end{figure}

A key advantage of the SAM framework is its ability to enable scalable design of ASMMs. To quantitatively evaluate this capability, we investigate a sinusoidal beam benchmark. The left end of the beam is fixed, and the target deformation requires its bottom edge to conform to a sinusoidal profile (Figs.~\ref{fig:scalability}a--i, blue dots). We systematically analyze design performance and accuracy under varying structural resolutions and deformation amplitudes. Resolution is varied by increasing the number of unit cells along both the vertical and horizontal directions, progressing from left to right in Figs.~\ref{fig:scalability}a--i. Deformation amplitude is varied across three levels (0.5, 1.0, and 1.5 times the beam width), progressing from top to bottom. For each combination of resolution and amplitude, results are presented for both microscale design methods, cDM and AS.

To quantify design efficiency, we evaluate the wall-clock time of the SAM framework under identical hardware configurations for both microscale design methods. Details of the computational setup are provided in Appendix~\ref{app:computational-setup}. With increasing resolution, both the cDM-based microscale design stage and the ConLME-based macroscale design stage exhibit linear scaling with respect to the total number of unit cells (Fig.~\ref{fig:scalability}j). A detailed time complexity analysis confirming this linear scaling, along with a comparison against baseline methods, is provided in Appendix~\ref{app:time-complexity}. In contrast, the computational cost of AS-based SAM grows superlinearly with resolution, approximately as $N_c^{2}$ between 180 and 1328 unit cells (Fig.~\ref{fig:scalability}m). As resolution increases, the runtime of AS-based SAM grows rapidly; at 1328 unit cells it is about 9 times that of cDM-based SAM (2961~s versus 322~s at the intermediate amplitude). This divergence arises from a fundamental difference in how the two methods scale: AS-based SAM resolves the macroscale subproblem after each search step to update the remaining free vertex positions and avoid error accumulation, a cost that grows substantially with the number of unit cells. The cDM-based SAM, by contrast, requires a full inverse diffusion process for each unit cell regardless of the total number, so its per-cell cost remains constant and the total cost grows only linearly.

Due to the lack of publicly available implementations of shape-morphing active metamaterials, we compare the time complexity of the SAM framework against a representative optimization-based baseline rather than against measured runtimes. Since gradient-based methods are generally more efficient than heuristic methods such as genetic algorithms, we take topology optimization (TO) as the reference. As derived in Appendix~\ref{app:time-complexity}, for a metamaterial structure composed of $N_c$ unit cells, each discretized by an $m \times m$ finite element grid, TO requires one finite element analysis of the assembled structure per iteration. Its cost over $k$ iterations scales as $O(km^4N_c^2)$ with a banded direct solver~\cite{golub_matrix_2013}, as $O(km^3N_c^{1.5})$ with a sparse direct solver~\cite{george_nested_1973}, and as $O(km^2N_c)$ in the best case of an optimal iterative solver~\cite{brandt_multi-level_1977, amir_multigrid-cg_2014}. The measured cost of cDM-based SAM grows linearly with $N_c$ (Fig.~\ref{fig:scalability}j). The advantage of SAM over TO therefore depends on the solver assumed for TO: with direct solvers it lies in the power of $N_c$, whereas against an optimal iterative solver both scale linearly and the advantage is the factor $km^2$, which arises because SAM requires no finite element analysis of the assembled structure during design. This comparison is asymptotic; it excludes the one-time offline cost of generating the dataset and training the model, and it is not an accuracy-matched runtime benchmark.

\subsection{Accuracy variation at two scales}\label{sec:scalability-accuracy}
To separate the shape-morphing performance at the macroscale and the microscale, we evaluate accuracy at each scale independently. At the microscale, we compute the RMS error $\mathrm{RMS}_{\mathrm{micro}}$ between the angle changes of the designed unit cells, simulated in isolation, and their target configurations. At the macroscale, we measure the mean absolute error (MAE) between the deformed positions of handle vertices and their prescribed targets (Figs.~\ref{fig:scalability}a--i). Calculation details for both metrics are provided in Appendix~\ref{sec:metrics}. Because the diffusion sampling is random, every cDM configuration is designed three times and the mean is reported; AS is deterministic.

At the microscale, the RMS error of unit cells generated by the cDM-based SAM is nearly constant across all resolutions (Fig.~\ref{fig:scalability}k); it remains close to the error measured on the test set (Appendix~\ref{app:cdm-model}), increases only slightly with the deformation amplitude, and varies little between repeated designs. The microscale error is therefore set by the generation accuracy of the diffusion model and does not grow with the number of unit cells. For AS-based SAM, the RMS error increases with deformation amplitude and likewise changes little with resolution (Fig.~\ref{fig:scalability}n). Because AS-based SAM retrieves pre-existing designs from the dataset and re-solves the macroscale subproblem after each retrieval, its unit cell targets coincide with dataset entries by construction, and its microscale error is consistently smaller than that of cDM-based SAM, by a factor that narrows as the amplitude increases.

At the macroscale, the MAE between the target handle vertex positions and those obtained from finite element simulations of the full structure is reported for both methods (Figs.~\ref{fig:scalability}l and \ref{fig:scalability}o). The MAE is normalized by the beam width. At the smallest amplitude, the coarse designs already have enough degrees of freedom to realize the required deformation, and the error remains small at all resolutions, with some fluctuation. At the largest amplitude, the limited degrees of freedom of the coarse designs cannot realize the large deformation, and the coarsest design has the largest error for both methods (Figs.~\ref{fig:scalability}a, d, and g); as the resolution increases, the additional degrees of freedom improve the deformation capability of the structure, and the error decreases. The two methods reach comparable errors at the two smaller amplitudes, whereas cDM-based SAM is more accurate at the largest. The repeated cDM designs differ noticeably only for the coarsest design at the smallest amplitude, where the spread between designs is comparable to the mean error; at higher resolutions the spread is small compared with the differences discussed here.

Although the unit cells retrieved by AS are more accurate than those generated by the cDM, the macroscale error of AS-based SAM is not smaller, and at the largest amplitude it is higher. The sources of these errors are analyzed in Section~\ref{sec:error-sources}, after the shape-morphing examples have been presented. Since the cost of cDM-based SAM grows linearly with the number of unit cells while that of AS-based SAM grows quadratically, and the macroscale accuracy of cDM-based SAM is comparable to or better than that of AS-based SAM, cDM-based SAM is used for the shape-morphing examples in Section~\ref{sec:generality}.

\begin{figure}[pos=!t]
    \centering
    \includegraphics[width=1.0\linewidth]{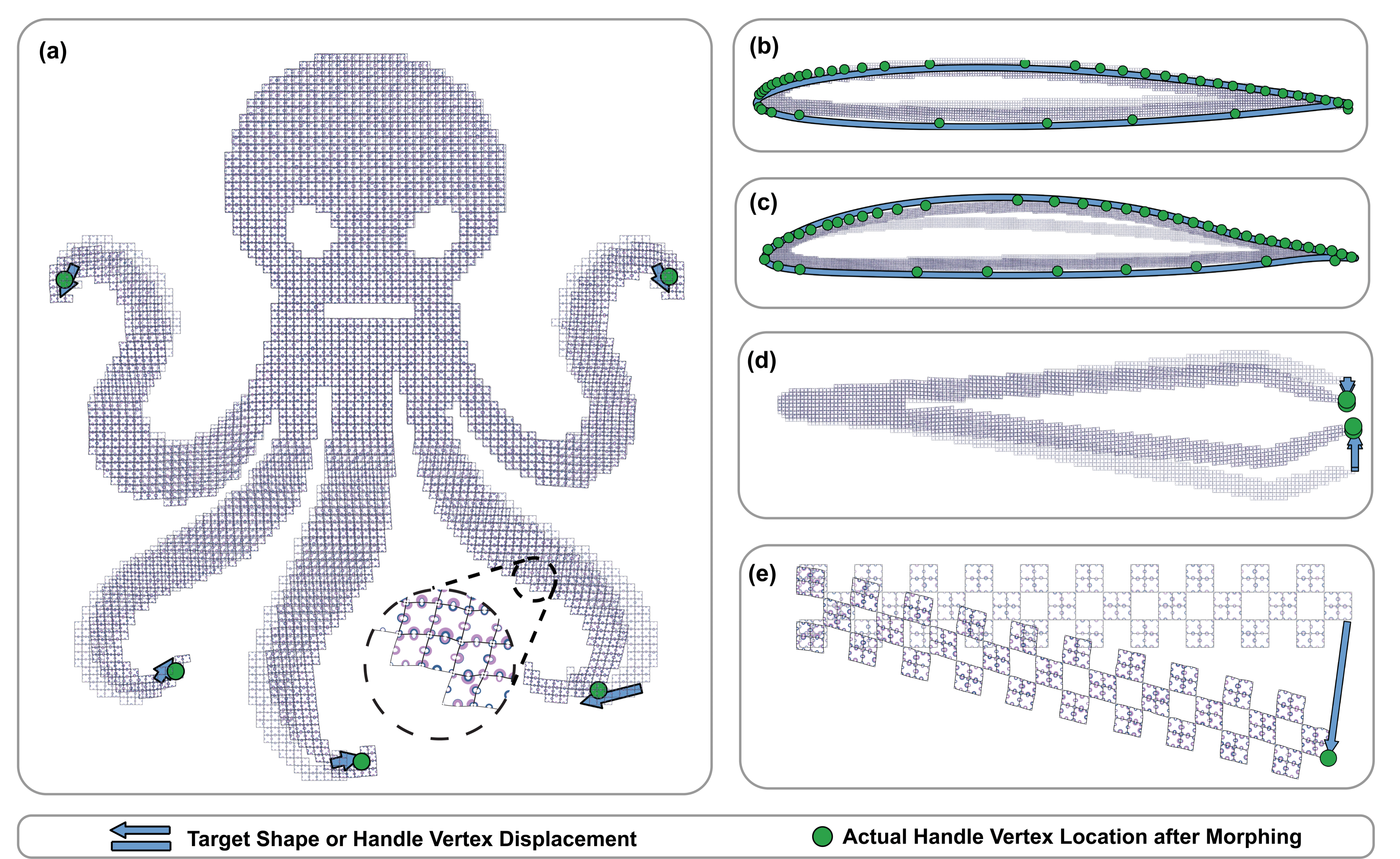}
    \caption{\textbf{Gallery of shape-morphing design examples generated by the proposed framework.} In all the figures, the green dot points out the actual deformation of the handle vertex after morphing, and the blue arrow presents the target displacement from the initial position to the target position. The faded structure shows the initial configuration.
    \textbf{a,} Octopus in which five arms are independently controlled to achieve different target deformations; the inset enlarges the designed unit cells of one arm.
    \textbf{b and c,} Two airfoil shape-matching tasks, with target shapes selected from the UIUC airfoil dataset \cite{selig_uiuc_1996, chen_airfoil_2020}. 
    Here, the target configuration is represented by a closed blue curve.
    \textbf{d,} Active metamaterial tweezers with the left end fixed and the two tips on the right approaching each other to achieve a gripping motion.
    \textbf{e,} Checkerboard beam structure demonstrating large deformation capability when sparse local units are introduced.}
    \label{fig:gallery}
\end{figure}

\subsection{Generality and shape-morphing accuracy}\label{sec:generality}
To demonstrate the generality of the proposed method, we designed several shape-morphing examples using the cDM-based SAM framework. In the first example, an octopus-shaped structure is actuated by assigning displacements of varying magnitudes and directions to the tip of each arm (Fig.~\ref{fig:gallery}a). Most arms approach their target positions, with the largest deviation at the lower-right arm, demonstrating the capability of the framework to handle multi-point actuation in geometrically complex structures. We further selected two airfoil geometries from the UIUC airfoil dataset \cite{selig_uiuc_1996, chen_airfoil_2020}, treating one as the initial configuration and the other as the target, and applied the SAM framework to design the morphing transition. The majority of handle vertices lie close to their target positions, with misalignment observed in several local regions (Figs.~\ref{fig:gallery}b and c). As a third example, we designed an active tweezer structure in which the left end is fixed while the two right-side tips converge to achieve a gripping function (Fig.~\ref{fig:gallery}d). Finally, when the domain is discretized into a checkerboard pattern---analogous to kirigami-inspired architectures---significantly larger deformations become achievable (Fig.~\ref{fig:gallery}e). Additional examples are provided in Appendix~\ref{app:examples}.

Two aspects of the problem setup matter for these examples. First, the linkage formed by the stiff members can bend but hardly elongate, so a handle target that lies farther from the fixed vertices than the handle's initial position cannot be reached; a purely vertical displacement of the tip of a cantilever is such a target. In the airfoil tail drop, octopus, and checkerboard beam examples, each handle target that would require the structure to elongate is therefore moved back, along the line to the centroid of the fixed vertices, to the initial distance from that centroid, while targets that bring the handle closer to the fixed vertices are kept as prescribed. Second, the airfoils are slender and their prescribed displacements are small compared with the chord, so that a small rigid rotation about the support dominates the error if the support is confined to the leading edge. The airfoils are therefore supported at a leading-edge vertex and at a distant handle vertex. The same supports are used in the macroscale solution and in the FEA.

Quantitative performance across the examples in Fig.~\ref{fig:gallery} is summarized in Table~\ref{tab:performance}, where design accuracy is evaluated using mean absolute error (MAE) and mean relative error (MRE), supplemented by the average coefficient of determination $R^2_\mathrm{macro}$. Definitions and calculation details of these metrics are provided in Appendix~\ref{sec:metrics}. The characteristic length $L_c$, defined as the maximum target displacement among all handle vertices, serves as the normalization reference for MRE. When only a single handle vertex is prescribed, as in the airfoil tail drop (Fig.~\ref{fig:overview}f), and checkerboard beam (Fig.~\ref{fig:gallery}e) examples, the MRE remains below 9\%. When multiple handle vertices are involved, as in the octopus (Fig.~\ref{fig:gallery}a) and tweezer (Fig.~\ref{fig:gallery}d) examples, or when the task requires matching an entire target shape, as in the airfoil morphing examples (Figs.~\ref{fig:gallery}b--c), the errors are larger. Since $R^2_\mathrm{macro}$ is computed from the handle positions rather than from their displacements, it stays close to one whenever the displacements are small compared with the size of the structure; MAE and MRE, which are based on displacements, are therefore the primary measures of accuracy.

\begin{table}[pos=!ht]
\centering
\caption{\textbf{Performance metrics of the example case studies.}}
\label{tab:performance}
\begin{tabular}{l c c c c c c c c c}
\toprule
Case & Deformation Plot&$N_c$ & $N_h$& $\mathrm{MAE}$ (mm) & $R^2_\mathrm{macro}$& $L_{\mathrm{c}}$ (mm) & $\mathrm{MRE}$\\
\midrule
Airfoil tail drop      & Fig.~\ref{fig:overview}f&2212 & 1 & 0.7598 & -     & 9.16 & 8.29\%\\
Octopus                & Fig.~\ref{fig:gallery}a&2950 & 5 & 2.0368 & 0.9971& 8.49 & 23.98\%\\
Airfoil shape change 1 & Fig.~\ref{fig:gallery}b&2209 & 43& 1.4734 & 0.9644& 5.77 & 25.51\%\\
Airfoil shape change 2 & Fig.~\ref{fig:gallery}c&4208 & 44& 1.5677 & 0.9948& 12.42& 12.62\%\\
Tweezers               & Fig.~\ref{fig:gallery}d&1673 & 4 & 1.0073 & 0.9237& 10.30& 9.78\%\\
Checkerboard beam      & Fig.~\ref{fig:gallery}e&120  & 1 & 0.2629 & -     & 10.05& 2.62\%\\
\bottomrule
\end{tabular}
\end{table}

\subsection{Error source analysis}\label{sec:error-sources}

The quantity that ultimately indicates whether a design succeeds is the macroscale error, i.e., how closely the assembled structure reaches its target shape. The macroscale error has two contributions. The first is the error of the unit cell designs, measured by the microscale error. The second is the assembly effect, i.e., the change in the deformation of a unit cell when it is embedded in the assembled structure. We discuss them in turn.

The error of the unit cell designs (microscale error) has two origins, namely the accuracy of the microscale design method and the feasibility of the targets required by the macroscale solution. Both can be identified from the macroscale solution with two diagnostic metrics, \textit{dissimilarity} and \textit{edge length deviation}. Dissimilarity measures the mean Euclidean distance between the vertices of each unit cell and those of its nearest neighbor in the dataset $\mathcal{D}$, with values approaching zero indicating closer agreement (Eq.~\eqref{Eq:dissimilarity} in Appendix~\ref{sec:weight-selection}). Edge length deviation quantifies how much each mesh edge deviates from its initial length $L_0 = 0.5$ mm, which preferably undergoes minimal variation throughout deformation. A high dissimilarity marks a unit cell whose required target configuration lies far from every entry of the dataset, i.e., an out-of-distribution (OOD) configuration, for which the design method may yield low accuracy. A large edge length deviation marks a unit cell required to stretch or shorten its edges significantly, which is beyond the capability of the stiff frame members. This problem arises where the prescribed handle positions cannot be met by an admissible deformation, i.e., when the targets are infeasible. These two quantities also serve as a posteriori admissibility checks on the macroscale solution. Because the consistency and target terms in Eq.~\eqref{Eq:Optimization problem} are soft penalties, ConLME does not guarantee feasible edge-length variation or configurations within the dataset, and cells with a high dissimilarity or a large edge-length deviation can indicate where the solution should not be trusted. With these metrics, the following discussion analyzes three origins of design error defined above---error from the microscale design method, error due to target infeasibility, and error resulting from the assembly effect---based on our design case studies.

\begin{figure}[pos=!htbp]
    \centering
    \includegraphics[width=0.9\linewidth]{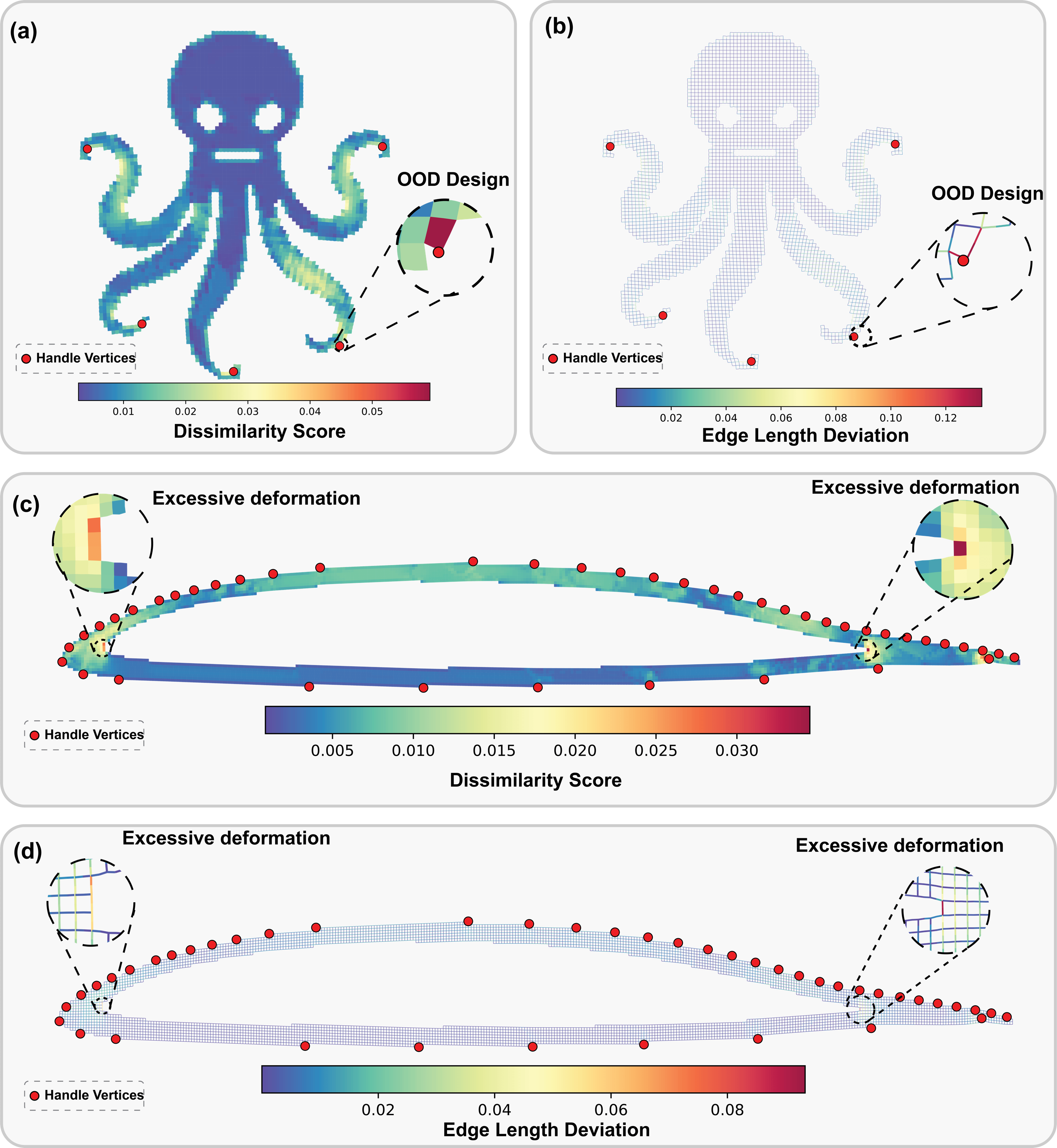}
    \caption{\textbf{Analysis of design error sources.} Dissimilarity score is measured at the unit-cell level as the mean Euclidean distance between the vertices of a cell and those of its closest counterpart in the dataset after coordinate alignment (Eq.~\eqref{Eq:dissimilarity}). Edge length deviation denotes the departure of each mesh edge length from its initial value $L_0 = 0.5$ mm. Both quantities are computed from the deformed mesh solved by ConLME.
    \textbf{a and b,} Dissimilarity score and edge length deviation of the octopus case shown in Fig.~\ref{fig:gallery}a. The zoom-in views highlight the out-of-distribution (OOD) design at the handle vertex of the lower-right arm.
    \textbf{c and d,} Dissimilarity score and edge length deviation of the airfoil shape-matching case shown in Fig.~\ref{fig:gallery}c. The zoom-in views highlight the excessively deformed cells (\textbf{c}) and edges (\textbf{d}) near the leading and trailing edges, at handle vertices whose prescribed positions the neighboring cells cannot follow.}
    \label{fig:error-sources}
\end{figure}

The microscale design method has an intrinsic error even for configurations covered by the dataset, and this error grows when the macroscale solution requires configurations that lie far from any entry in $\mathcal{D}$. For example, the unit cell configurations around the handle vertex at the tip of the lower-right arm of the octopus structure exhibit the highest dissimilarity (Fig.~\ref{fig:error-sources}a) and the largest edge length deviation (Fig.~\ref{fig:error-sources}b) of the structure. These configurations represent OOD targets and substantially increase the difficulty of local inverse design, and the resulting infills may produce deformation responses that deviate significantly from the target. The arm that contains these cells is also the one with the largest deviation from its target (Fig.~\ref{fig:gallery}a).

The second error origin is the feasibility of the targets. When the prescribed handle vertex positions cannot be met with unit cell configurations close to the dataset, the optimization, whose constraints are soft, produces a compromised solution in which some unit cells are forced into excessive deformation. This feasibility challenge is further compounded as the number of handle vertices increases, since their prescribed positions impose mutual constraints on one another, making it increasingly difficult to identify a globally consistent target configuration. This is evident in the airfoil shape-matching task of Fig.~\ref{fig:gallery}c. Near the leading edge and near the thin trailing edge, the unit cells around the handle vertices are forced into excessive deformation, with the highest dissimilarity and edge length deviation of the structure (Figs.~\ref{fig:error-sources}c and \ref{fig:error-sources}d), indicating that the targets there are not reached with configurations covered by the dataset. Because the two metrics are related, they do not fully separate the possible causes, namely a target that is physically unreachable, a region of the configuration space that the dataset does not cover, or a macroscale solution that is not the best one available. This error can be mitigated by relaxing the target constraints on the handle vertices. For example, instead of prescribing exact target positions, we can require the handle vertices to lie on the target curve to allow greater flexibility in the global optimization.

Once the unit cell designs are accurate enough, the assembly effect dominates the macroscale error. This is the case in the sinusoidal beam benchmark (Section~\ref{sec:scalability-accuracy}). The smaller microscale error of AS-based SAM does not translate into a smaller macroscale error (Figs.~\ref{fig:scalability}k, \ref{fig:scalability}l, \ref{fig:scalability}n, and \ref{fig:scalability}o), especially at the largest deformation amplitude, where the macroscale error of AS-based SAM is higher than that of cDM-based SAM (Figs.~\ref{fig:scalability}l and \ref{fig:scalability}o). The assembly effect originates from the hinges. The framework treats the stiff members as a linkage of rigid bars connected by ideal hinges, which transmit no moment, so that neighboring unit cells do not interact. The compliant hinges used in practice have a finite rotational stiffness and transmit moments between neighboring unit cells. The dataset is generated from isolated unit cells and does not contain this interaction. In the assembled structure, neighboring cells share their frame members and hinges, and each cell deforms more than its isolated counterpart (Appendix~\ref{app:assembly}). The stiff--soft discretization thus reduces, but does not eliminate, the interaction between unit cells. The microscale error does not capture the assembly effect since it is evaluated on isolated unit cells. The two microscale design methods, AS and cDM, differ in how their microscale errors combine with the assembly effect. The unit cells retrieved by AS more closely match the required deformation for isolated cells, but this leads to excess deformation in the assembled structure due to additional deformation caused by interaction between cells. In contrast, the unit cells generated by the cDM may struggle to reproduce extremely large deformations because such deformation regimes are sparsely represented in the training data (Appendix~\ref{app:assembly}). Consequently, at the largest deformation amplitude, the cDM-based design attains a lower macroscale error, although its unit cells are less accurate at the microscale. This may happen because the deformation due to cell interactions and the effect of insufficient deformation at the single-cell level are coincidentally compensated. The difference between the two methods at the macroscale thus reflects how their microscale errors combine with the assembly effect.

\subsection{Compactness}
Another property of the SAM framework concerns the area occupied by the structure. For an idealized mechanism of rigid bars and densely arranged cells, the area of the region spanned by the corner vertices of the unit cells can only decrease upon deformation, regardless of the specific deformation mode, as proved in Appendix~\ref{app:volume}. Because the structures considered in this work are two-dimensional, volume here refers to this in-plane area. The result bounds the area of this idealized region; it does not bound the outer envelope of the finite structure, whose boundary hinges may bulge outward, nor the space swept during actuation.

\begin{figure}[pos=!htbp]
    \centering
    \includegraphics[width=0.9\linewidth]{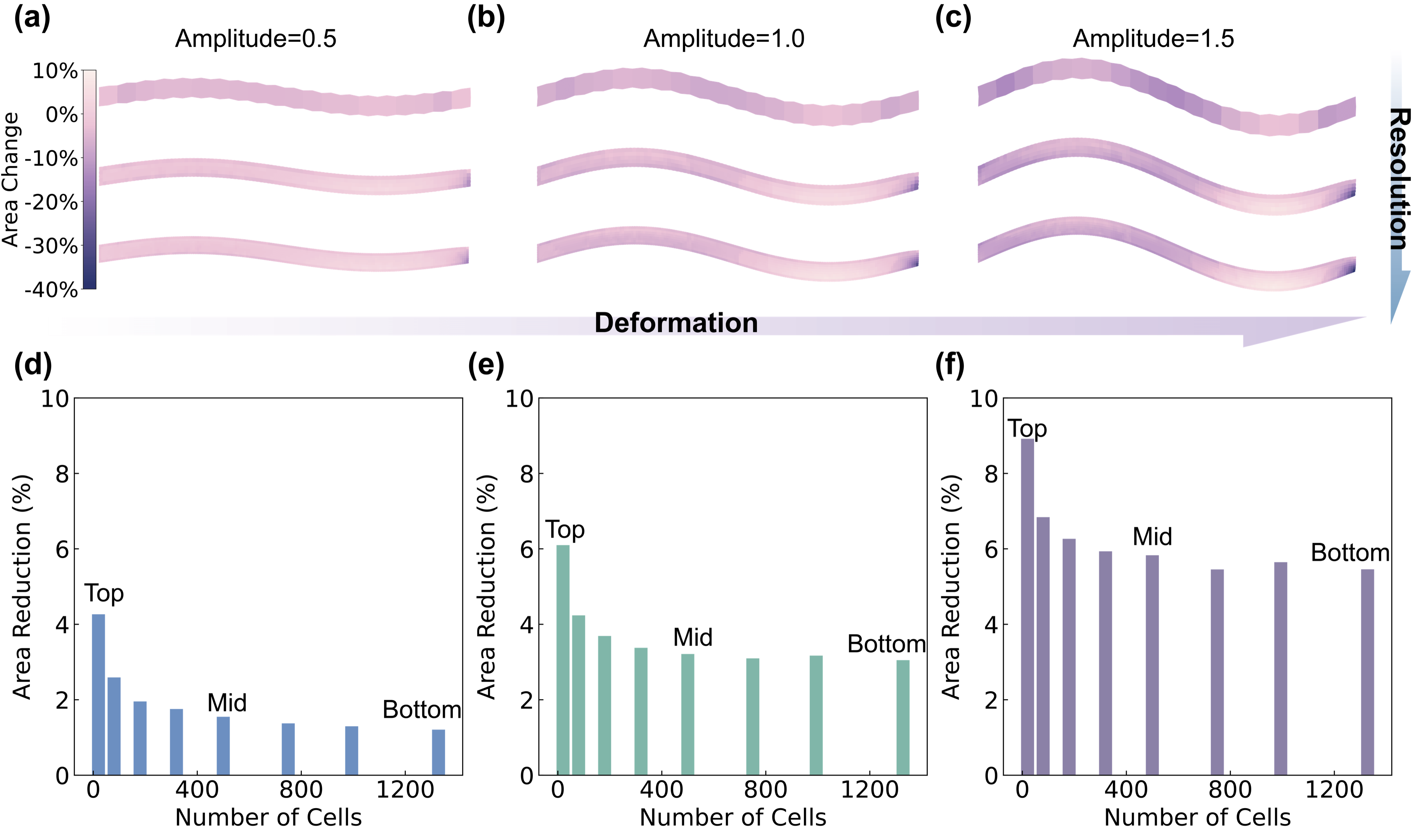}
    \caption{\textbf{Volume change analysis of structures designed using the SAM framework.}
    \textbf{a--c,} Spatial distribution of the area change of the unit cells in cDM-based SAM designs for deformation amplitudes of 0.5 (\textbf{a}), 1.0 (\textbf{b}), and 1.5 (\textbf{c}) times the beam width. In each panel, the resolution increases from top to bottom (20, 500, and 1328 unit cells). Each cell is measured by the quadrilateral spanned by its four corner vertices, and all maps share the color scale on the left.
    \textbf{d--f,} Total volume change, reported as the area reduction of the region $\Omega$ (Appendix~\ref{app:volume}), as a function of the number of unit cells for deformation amplitudes of 0.5 (\textbf{d}), 1.0 (\textbf{e}), and 1.5 (\textbf{f}). The labels Top, Mid, and Bottom mark the resolutions shown in the top, middle, and bottom rows of \textbf{a--c}.}

    \label{fig:volume}
\end{figure}

To numerically verify this property, we examine the distribution of local volume changes across the structure following the macroscale solution for three deformation amplitudes (Figs.~\ref{fig:volume}a--c). The maximum local volume change increases with deformation amplitude. We further evaluate the total volume change of the region $\Omega$ defined in Appendix~\ref{app:volume}, i.e., the union of the quadrilaterals spanned by the four corner vertices of each unit cell. Measuring $\Omega$ excludes the boundary effect neglected in the proof: the hinge vertices on the outer boundary may bulge outward, whereas inside the domain such bulges cancel between neighboring cells. Since the distances between adjacent vertices generally decrease upon deformation, the area of $\Omega$ decreases in all cases, as predicted. For the smallest deformation amplitude, the reduction is small and decreases from 2.6\% to 1.2\% as the resolution increases from 80 to 1328 cells (Fig.~\ref{fig:volume}d); at larger amplitudes it becomes more pronounced but remains below 7\% (Figs.~\ref{fig:volume}e--f). The coarsest design (20 cells) shows larger reductions, of up to 8.9\%. When the outer boundary is included, the total area deviates from that of $\Omega$ by 6--7\% for 20 cells and by less than 1\% for 1328 cells, confirming that the boundary effect vanishes with increasing resolution. Collectively, these results show that, in this benchmark, the area of $\Omega$ changes by less than 7\% for designs with at least 80 unit cells.

\section{Discussion}\label{sec:discussion}
Scalable design frameworks are essential for shape-morphing metamaterials as deformation tasks grow in geometric complexity and structural resolution. Conventional approaches based on global optimization rely heavily on physics-based simulations and iterative updates, rendering them computationally impractical at high resolutions. Data-driven inverse design methods, while capable of accelerating individual design queries, are often tailored to specific problem configurations and thus lack the generality required for broad deployment. In contrast, the SAM framework exploits the complementary mechanical properties of soft and stiff materials to approximately decouple macroscale and microscale displacements, yielding a framework that is simultaneously general, efficient, and scalable. 

The unit-cell-based discretization in SAM offers three key advantages over bar-based alternatives. First, the interactions among the members of a unit cell, which a bar-based model neglects, are captured through finite element simulations of complete unit cells during dataset generation. Second, each unit cell encodes a unique directional deformation mode via geometric asymmetry, eliminating the ambiguity inherent in axially symmetric bar elements. Finally, the two-dimensional infill domain provides a richer design space, allowing both material parameters and structural topology to be tuned while accommodating functional components such as sensors and actuators. 

We construct a unit-cell-level dataset of infill geometries and their corresponding deformation responses, which serves dual purposes: constraining the macroscale solution and training microscale inverse design models. At the macroscale, the ConLME method efficiently resolves global mechanism deformation under data-driven constraints, producing target configurations for subsequent microscale inverse design. At the microscale, two alternative inverse design methods determine the final infill geometry for each unit cell. Crucially, because the dataset is constructed at the unit-cell level, it remains reusable across arbitrary macroscale design domains---a key distinction from problem-specific data-driven approaches that require retraining for each new configuration.

The SAM framework is validated through geometrically nonlinear finite element simulations across a range of representative examples, demonstrating its generality in handling diverse deformation tasks. When combined with the conditional diffusion model, the computational cost scales approximately linearly with the number of unit cells, confirming the scalability of the framework. Furthermore, unlike kirigami-based designs that typically involve substantial volume changes during actuation, the area spanned by the unit cells of a SAM structure does not increase upon deformation, a property that may be of interest for space-constrained applications, although the occupied envelope and the swept space during actuation are not bounded by this result.

While thermally activated materials are adopted here as a representative example, the SAM framework is not inherently tied to a specific actuation mechanism and can be readily extended to other material systems---such as magnetoactive or electroactive materials---provided that a corresponding unit cell dataset is available. Furthermore, while this work employs a compact parameterized representation for unit cell infill geometries, the framework is compatible with more general free-form geometric representations. Such a substitution would not alter the overall framework architecture, but would impose greater demands on the capacity and generalization ability of the microscale inverse design model.

While the SAM framework has demonstrated its potential as a scalable and general approach to active metamaterial design, it remains in an early stage of development, and several limitations warrant further investigation. First, the decoupling of the two scales is approximate. The dataset describes isolated unit cells, whereas the compliant hinges of an assembled structure transmit moments between neighboring unit cells, so that a unit cell deforms differently inside the assembled structure than in isolation (Section~\ref{sec:error-sources}). Generating the dataset from unit cells embedded in a representative neighborhood is a possible remedy. In addition, the verification model does not include contact between the infill members, and its sensitivity to the finite element mesh has not been studied. Second, since the macroscale formulation is adapted from Laplacian mesh editing, it inherits a tendency to produce local distortions under extremely large deformations. A potential remedy is to replace the Laplacian mesh editing initialization with the large-deformation-oriented as-rigid-as-possible (ARAP) method \cite{igarashi_as-rigid-as-possible_2005}, at the cost of increased computational time. Third, the placement of handle vertices and their prescribed displacements significantly influences the design outcome, and inappropriate configurations may lead to violations of mesh topology; in particular, targets that require the structure to elongate cannot be reached (Section~\ref{sec:generality}), and the accuracy of slender structures is sensitive to where they are supported. Furthermore, the soft constraint weights $w_L$, $w_S$, and $w_t$ are currently tuned manually through the iterative procedure described in Appendix~\ref{sec:weight-selection}. Developing a fully automated weight selection strategy remains an open problem. Fourth, the use of a regular mesh at the macroscale limits the smoothness of the structural boundary, which could be addressed by adopting a more flexible domain discretization, such as a non-uniform triangular mesh. Finally, the coverage of the unit cell dataset constrains the achievable range of local deformations; this limitation could be mitigated by incorporating materials or structures capable of producing larger deformations into the dataset, thereby expanding the accessible design space. Addressing these limitations in future work will further extend the applicability of the SAM framework toward fully autonomous, high-fidelity design of active metamaterial systems.
\section{Conclusions}\label{sec:conclusions}
This work presents the scalable active metamaterial (SAM) framework for designing aperiodic shape-morphing active metamaterials. A hybrid stiff--soft unit cell discretization approximately decouples the design problem into two scales that can be solved without simulating the assembled structure. The macroscale subproblem is solved by constrained Laplacian mesh editing (ConLME) with soft data-driven feasibility constraints. In the microscale subproblem, the infill geometry of each unit cell is inversely designed using either a conditional diffusion model (cDM) or an adjustable search (AS) strategy. Both subproblems rely on a unit-cell-level dataset that is reusable across different design problems.

With the cDM, the design cost of SAM grows linearly with the number of unit cells. The framework applies to arbitrary initial geometries. The same dataset and trained model were used to design an octopus with independently driven arms, morphing airfoils, tweezers, and a checkerboard beam. We employed geometrically nonlinear finite element simulations of the assembled structures to validate the final solution. Results show that the mean relative errors measuring the deviation of the actual deformed shapes from the targets are between 3\% and 29\%. The error stems from the accuracy of the unit cell designs, the infeasibility of the target, and the residual interaction between neighboring unit cells, which the stiff--soft discretization reduces but does not eliminate. In addition, for an idealized rigid-bar mechanism, the area spanned by the corner vertices of the unit cells can only decrease upon deformation, a property of interest for space-constrained applications. Together, these results show that SAM provides a scalable route to the design of highly complex shape-morphing active metamaterials. While we demonstrated this framework under a specific infill representation and actuation mechanism setting, the idea of decoupling the design problem into two scales and the general design methodology at each scale can extend beyond the current form.

\section*{Data availability}
The dataset required to interpret, verify, and extend the results of this study is available via Hugging Face at \url{https://huggingface.co/datasets/digitlab/SAM-Dataset}. The pre-trained conditional diffusion model is also available at: \url{https://huggingface.co/digitlab/SAM-Diffusion}.


\section*{Acknowledgments}
This work was supported by the startup funds from the J. Mike Walker ’66 Department of Mechanical Engineering at Texas A\&M University. Portions of this research were conducted with the advanced computing resources provided by Texas A\&M High Performance Research Computing.

\clearpage
\appendix
\setcounter{equation}{0}
\setcounter{figure}{0}
\setcounter{table}{0}
\renewcommand{\theequation}{S\arabic{equation}}
\renewcommand{\thefigure}{S\arabic{figure}}
\renewcommand{\thetable}{S\arabic{table}}
\renewcommand{\theHequation}{supp.\arabic{equation}}
\renewcommand{\theHfigure}{supp.\arabic{figure}}
\renewcommand{\theHtable}{supp.\arabic{table}}
\makeatletter
\renewcommand\@seccntformat[1]{\ifstrequal{#1}{section}{Appendix~}{}\csname the#1\endcsname\@seccntDot\hskip 0.5em}
\makeatother
\section{Unit cell design and geometric parameters}\label{sec:unit-cell}\label{app:unit-cell-params}
\begin{figure}[pos=!htbp]
    \centering
    \includegraphics[width=0.75\linewidth]{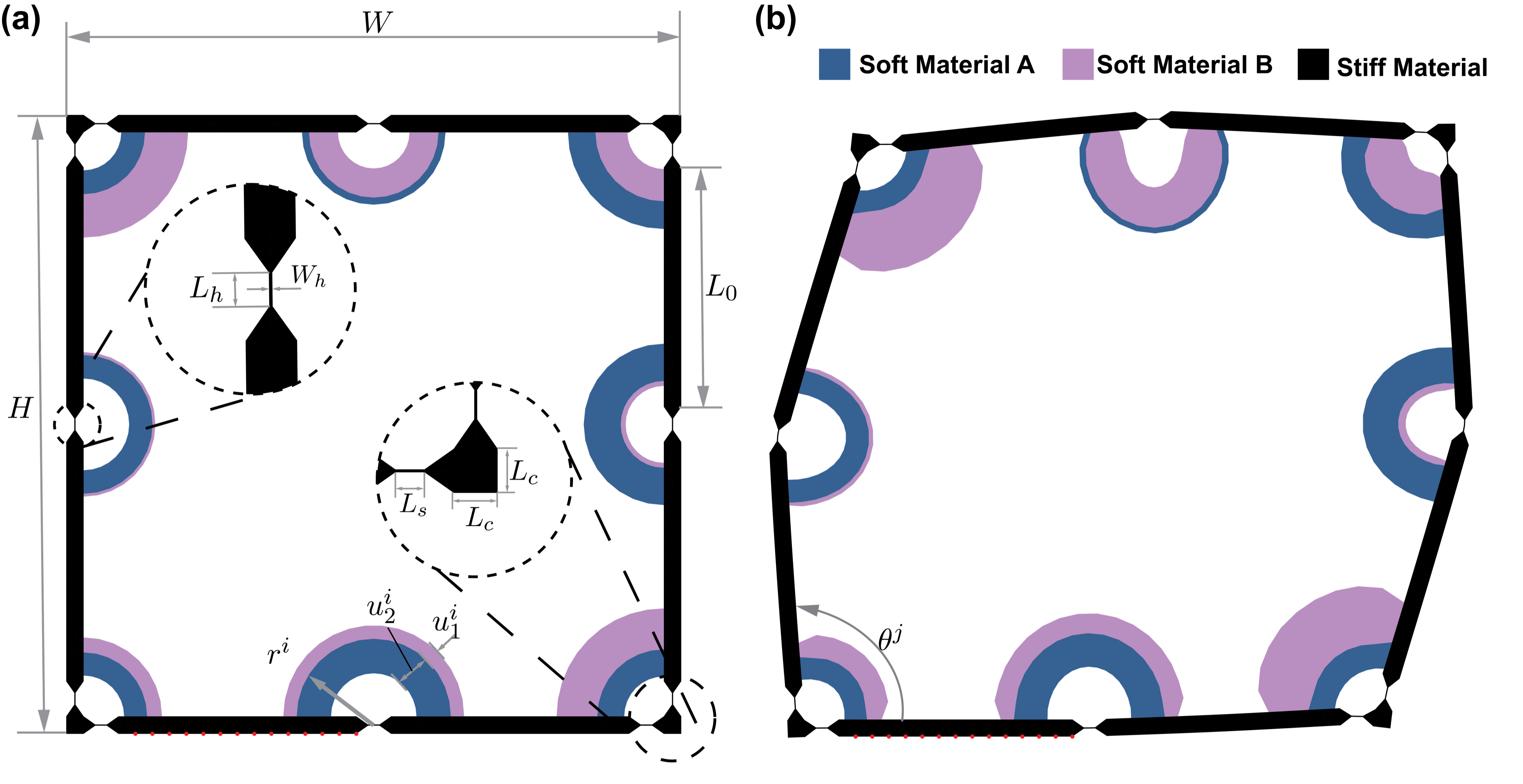}
    \caption{\textbf{Unit-cell geometry and deformation configuration.}
    \textbf{a,} Geometric parameterization of the CRL and its infill structure.
    \textbf{b,} The resulting deformed configuration.}
    \label{fig:unit-cell}
\end{figure}

The geometric configuration of the unit cell is illustrated in Fig.~\ref{fig:unit-cell}a. 
The unit cell is defined over a rectangular domain $\Omega = [0,W] \times [0,H]$. 
For simplicity, we consider the square case with $W = H$. 
The unit cell consists of two components: a closed-loop rigid linkage (CRL), shown in black, and an internal soft infill composed of stacked arc-shaped beams \cite{zhang_temperature-responsive_2024, zhang_shape_2022}.

\subsection{Geometric parameterization of the CRL}
As shown in Fig.~\ref{fig:unit-cell}a, the CRL is formed by stiff material. 
Each side of the CRL has a length $L$ and consists of two rigid bars, each of length $L_0 = L/2$ and width $L_c$. 
Adjacent bars are connected by a neck-shaped short link of length $L_s$, whose narrowest section has a width $W_h$ and a length $L_h$. 
This geometry localizes bending at the connection and guides deformation to occur in the hinge region. The values or ranges of all parameters are shown in Table~\ref{tab:unit-cell-params}.

\subsection{Geometric parameterization of the soft infill structure}
The internal infill consists of eight stacked circular arc beams arranged around the eight hinge centers, with each beam characterized by distinct geometric parameters. 
Each stacking arc beam $i$ is composed of two layers of materials with different coefficients of thermal expansion; the thicknesses of the outer and inner layers are denoted by $u_1^i$ and $u_2^i$, respectively. The inner radius of the inner layer is denoted by $r^i$, so that the layer interface and the outer surface of the beam lie at radii $r^i + u_2^i$ and $r^i + u_2^i + u_1^i$, respectively.
The two layers are made of an active material A with a large coefficient of thermal expansion and a passive material B with a negligible one (Appendix~\ref{app:fem}). To allow each beam to bend either inward or outward, an additional binary parameter $z^i$ is introduced to indicate the stacking order, where $z^i = 1$ assigns the active material A to the inner layer and the passive material B to the outer layer, and $z^i = 0$ indicates the opposite arrangement.
Consequently, each stacked arc beam is parameterized by four variables, resulting in a total of 32 geometric parameters for a single unit cell, i.e., $\boldsymbol{x}=[r^0,u_1^0,u_2^0,z^0,\ldots,r^7,u_1^7,u_2^7,z^7]$. The range of parameters is shown in Table~\ref{tab:unit-cell-params}.

Fig.~\ref{fig:unit-cell}b illustrates the deformation configuration of the ideal CRL, in which the bars are connected by perfect hinges. 
Since all CRL bars have equal length, the deformation configuration can be fully characterized by the variations of the eight internal angles $\boldsymbol{\theta}=[\theta^0,\ldots,\theta^7]$.

\begin{table}[pos=!htbp]
    \centering
    \caption{\textbf{Geometric parameters of a unit cell.}}
    \begin{tabular}{ccc}
        \toprule
        Parameter & Value (mm) & Description \\
        \midrule
        $L$   & 1.00          & Length of the infill region ($L = 2L_0$) \\
        $L_0$ & 0.50          & Length of a rigid bar \\
        $L_c$ & 0.03          & Width of the CRL bar \\
        $L_s$ & 0.06          & Length of the neck-shaped link \\
        $W_h$ & 0.002         & Width of the narrowest section of the neck-shaped link \\
        $L_h$ & 0.02          & Length of the narrowest section of the neck-shaped link \\
        $H$   & 1.06          & Total height of the unit cell \\
        $W$   & 1.06          & Total width of the unit cell \\
        $r^i$ & [0.06, 0.08]  & Inner radius of the inner layer \\
        $u_1^i$ & [$10^{-3}$, 0.08] & Thickness of the outer layer \\
        $u_2^i$ & [$10^{-3}$, 0.08] & Thickness of the inner layer \\
        $z^i$   & \{0,1\}        & Stacking order\\
        \bottomrule
    \end{tabular}
    \label{tab:unit-cell-params}
\end{table}

\section{FEM setup and materials properties}\label{app:fem}
In this work, thermal activation is used to drive material deformation. With a thermal strain of 30\% in the active material and rotations of the unit-cell members of several tens of degrees, the deformations are not small, so the response is modeled with geometrically nonlinear, finite-strain thermoelasticity. The deformation gradient $\boldsymbol{F} = \boldsymbol{1} + \nabla\boldsymbol{u}$ is split multiplicatively into an elastic and a thermal part,
\begin{equation}
    \begin{aligned}
        \boldsymbol{F} = \boldsymbol{F}_e\boldsymbol{F}_{\theta}, \qquad \boldsymbol{F}_{\theta} = (1+\alpha\Delta T)\,\boldsymbol{1},
    \end{aligned}
\end{equation}
where $\alpha$ is the coefficient of thermal expansion (CTE), $\Delta T$ is the temperature change, and $\boldsymbol{1}$ is the second-order identity tensor. The elastic part is governed by a compressible neo-Hookean strain energy per unit volume of the thermally expanded configuration,
\begin{equation}
    \begin{aligned}
        \psi(\boldsymbol{F}_e) = \frac{\mu}{2}\left(\operatorname{tr}\boldsymbol{C}_e - 3 - 2\ln J_e\right) + \frac{\lambda}{2}\left(\ln J_e\right)^2, \qquad \boldsymbol{C}_e = \boldsymbol{F}_e^{\mathsf{T}}\boldsymbol{F}_e, \quad J_e = \det\boldsymbol{F}_e,
    \end{aligned}
    \label{Eq:neo-hookean}
\end{equation}
with the Lam\'e constants $\lambda$ and $\mu$. The displacement field makes the total energy $\int_{\Omega_0} \psi(\boldsymbol{F}_e)\,\det\boldsymbol{F}_{\theta}\,\mathrm{d}V$ stationary; no external forces are applied.

The CTE of the active material A is set to $0.003\,^\circ\mathrm{C}^{-1}$, with a temperature change of $\Delta T = 100^\circ\mathrm{C}$, yielding a sufficiently large deformation configuration space. The passive infill material B is assigned a negligible CTE of $5\times10^{-6}\,^\circ\mathrm{C}^{-1}$, and the stiff CRL material a CTE of $10^{-10}\,^\circ\mathrm{C}^{-1}$, to minimize their contributions to the thermal deformation. The two infill materials share the same Young's modulus $E$, while the CRL is assigned $1000E$; this stiffness ratio of 1000:1 realizes the assumption in Section~\ref{sec:framework} that the deformation of the stiff members is negligible. All materials have a Poisson's ratio of $\nu = 0.3$. Since the loading is purely thermal and only displacement boundary conditions are applied, the displacement field depends on the stiffness ratio but not on the absolute value of $E$, which is therefore normalized to $E = 1$. The problem is solved in two dimensions under the plane-strain assumption, i.e., the out-of-plane stretch of $\boldsymbol{F}$ is one, with $\lambda = E\nu/[(1+\nu)(1-2\nu)]$ and $\mu = E/[2(1+\nu)]$. Geometric modeling and mesh generation are performed using Gmsh \cite{geuzaine_gmsh_2009}, and the resulting mesh of linear triangular elements is imported into FEniCSx \cite{alnaes_unified_2014, scroggs_construction_2022, baratta_dolfinx_2023} for finite element analysis. The nonlinear problem is solved with Newton's method using PETSc \cite{petsc-efficient, petsc-web-page, petsc-user-ref, dalcinpazklercosimo2011}, with a sparse direct solver for the linearized systems and a relative residual tolerance of $10^{-6}$. The temperature change is applied incrementally, starting with four load steps; a step that fails to converge is halved and repeated, and a step that converges in at most five iterations enlarges the next one. Contact between the members of the infill is not modeled.

Boundary conditions differ between dataset generation and full-structure simulation. During dataset generation, the bottom-left bar of a unit cell (Appendix~\ref{sec:unit-cell}) is fixed, and the response is extracted as the changes of the eight interior angles of the unit cell between the undeformed and the deformed configuration, measured at the mesh nodes nearest to the eight hinge points. In full-structure simulations of the assembled metamaterial, the fixed vertices prescribed in the upstream design stage are enforced as displacement boundary conditions, with all degrees of freedom of the finite element containing each fixed vertex constrained.

\section{Data generation}\label{sec:data-generation}
Infill structures, parameterized as described in Appendix~\ref{sec:unit-cell}, are generated by uniformly sampling the parameter space, with lower and upper bounds imposed on beam thickness and radius (Table~\ref{tab:unit-cell-params}) to prevent interference between adjacent beams and ensure proper connectivity. Finite element analysis (FEA; Appendix~\ref{app:fem}) is then performed on each sampled design to obtain the corresponding deformation response. Of the 60,000 sampled designs, one did not converge in the nonlinear analysis and was discarded, yielding a dataset $\mathcal{D}$ of 59,999 samples. The response of each design is stored as the changes of the eight interior angles of the unit cell between the undeformed and the deformed configuration. No design leaves the unit cell undeformed: the smallest RMS angle change in the dataset is $1.7^\circ$. The dataset serves two purposes within the SAM framework. First, it provides the data-driven soft constraints in the macroscale subproblem (Eq.~\eqref{Eq:Optimization problem}), ensuring that the resulting unit cell configurations remain feasible for microscale inverse design. Second, it directly supports the microscale subproblem, either as training data for the conditional diffusion model or as a retrieval dataset for the AS-based approach.

\section{Weight selection and iterative convergence of ConLME}\label{sec:weight-selection}
Adjusting the weights of the three energy terms shifts the emphasis of the optimization. In practice, $w_t$ is set to a relatively large value to ensure that the handle vertices achieve their prescribed displacements, $w_S$ is increased progressively from a small initial value until the configuration dissimilarity reaches an acceptable level, and $w_L$ is set empirically. Therefore, the analysis mainly focuses on the relative magnitude of the first two terms. By controlling their ratio $r = w_S / w_L$, we evaluate its influence using two diagnostic metrics. \textit{Configuration dissimilarity} measures the deviation of each unit cell from its nearest neighbor in $\mathcal{D}$, indicating whether the configuration is out-of-distribution and thus whether it can be reliably handled by the subsequent microscale inverse design. \textit{Area change} quantifies the magnitude of local deformation within each unit cell, serving as an indicator of realization difficulty (i.e., a large area change may indicate that the deformation is physically unrealizable). The two metrics are defined below. The \textit{area change} is
\begin{equation}
    \begin{aligned}
        \Delta A= A'-A^0,
    \end{aligned}
\end{equation}
where $A^0 = L^2 = 1\,\mathrm{mm}^2$ is the initial area of each unit cell. The area of the deformed configuration is computed using the shoelace formula \cite{braden_surveyors_1986}:
\begin{equation}
    \begin{aligned}
        A'=\frac{1}{2}\left|\sum_{i=1}^8\left( x_iy_{i+1}-x_{i+1}y_i\right)\right|,
    \end{aligned}
\end{equation}
where $(x_i, y_i) = \boldsymbol{v}_i \in \mathcal{V}_\mathrm{unit}$ are the vertex coordinates of the unit cell, with the convention $\boldsymbol{v}_9 = \boldsymbol{v}_1$. The \textit{configuration dissimilarity} is computed as
\begin{equation}
    \begin{aligned}
        \Delta S = \frac{1}{8}\sum_{j=1}^8\left\| \boldsymbol{v}_j-\hat{\boldsymbol{v}}_j\right\|,
    \end{aligned}
    \label{Eq:dissimilarity}
\end{equation}
where $\boldsymbol{v}_j$ denotes the coordinate of the $j$th vertex in the deformed configuration, and $\hat{\boldsymbol{v}}_j$ denotes the coordinate of the $j$th vertex in the most similar configuration from the dataset after the optimal rigid alignment of Eq.~\eqref{Eq:retrieval}.

\begin{figure}[pos=!htbp]
    \centering
    \includegraphics[width=1.0\linewidth]{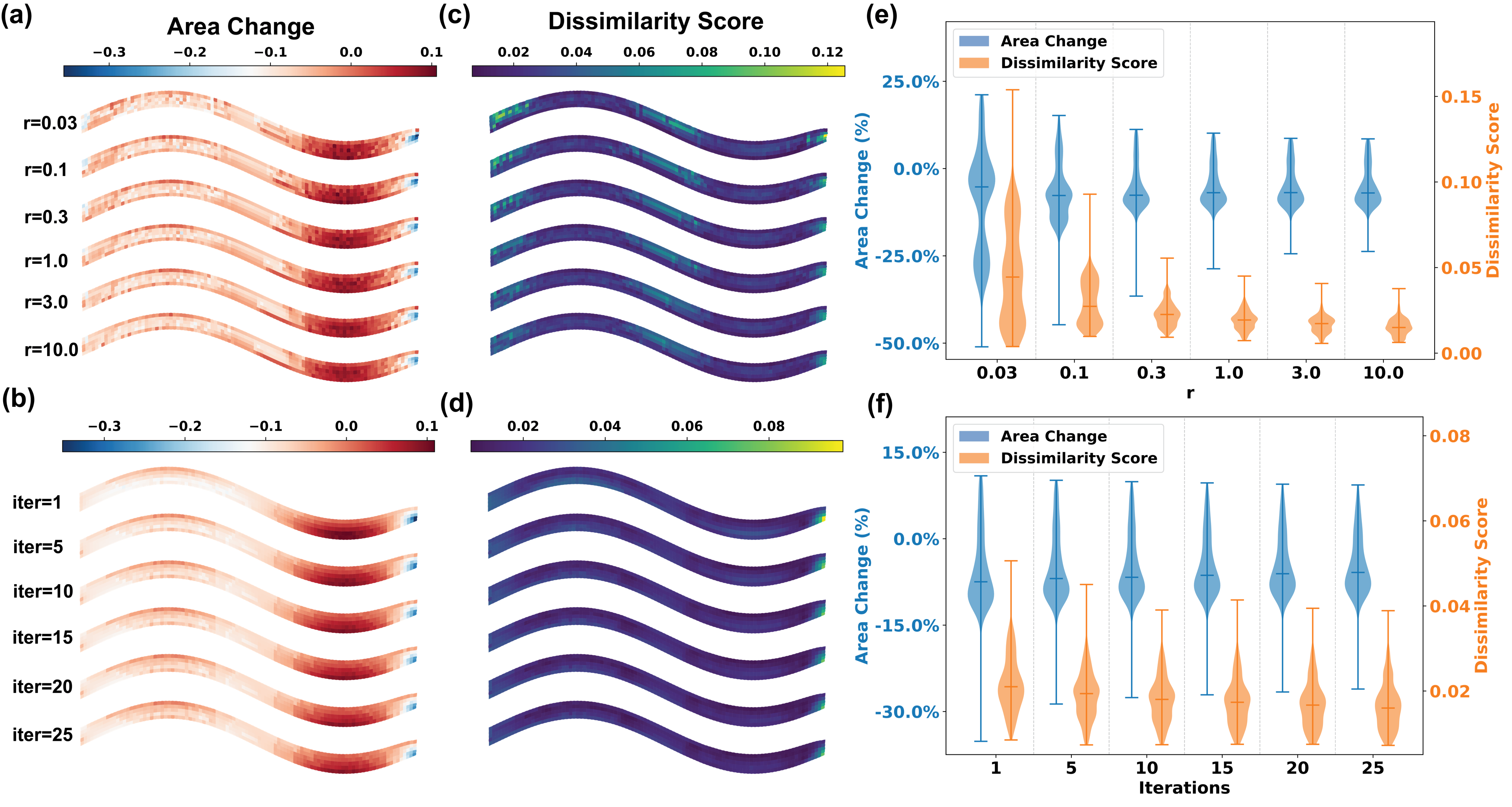}
    \caption{\textbf{Effect of weight ratio $r$ and iteration count on ConLME performance, illustrated using the sinusoidal beam benchmark.}
    \textbf{a, b,} Distributions of area change under varying weight ratios $r$ (\textbf{a}) and iteration counts (\textbf{b}).
    \textbf{c, d,} Distributions of configuration dissimilarity under varying weight ratios $r$ (\textbf{c}) and iteration counts (\textbf{d}).
    \textbf{e, f,} Evolution of the area change distribution and configuration dissimilarity score distribution with increasing $r$ (\textbf{e}) and iteration count (\textbf{f}).}
    \label{fig:conlme-weights}
\end{figure}

We use the sinusoidal beam benchmark (Section~\ref{sec:scalability}) to illustrate how the two metrics vary with the weight ratio $r$, varied through $w_S$ with $w_L = w_t = 1$ (Figs.~\ref{fig:conlme-weights}a and c). When $r$ is small, the configuration consistency energy is too weak to hold the unit cells near feasible configurations, and the Laplacian energy, which pulls every vertex toward the centroid of its neighbors, contracts the mesh: at $r = 0.03$, the area change spreads down to about $-50\%$ and the dissimilarity score reaches about 0.15, with the most distorted cells concentrated along the boundary (Fig.~\ref{fig:conlme-weights}e). As $r$ increases, the consistency energy pulls each unit cell toward its nearest configuration in $\mathcal{D}$, contracting the tails of both the area change and the dissimilarity distributions. For $r \geq 1$, both distributions remain essentially unchanged, indicating that the macroscale solution is insensitive to the precise value of $w_S$ once the consistency constraint is sufficiently strong. 

As the deformed configuration evolves during optimization, the nearest-neighbor configuration retrieved from the dataset, and hence the term $\boldsymbol{b}^{\mathcal{D}}$ in Eq.~\eqref{Linear system}, is updated accordingly, making the optimization problem itself iteration-dependent. Multiple iterations of the alternating scheme above are therefore required until the weighted energy converges. The influence of the iteration count is analyzed in Figs.~\ref{fig:conlme-weights}b and d. The distributions of area change and configuration dissimilarity as a function of iteration count are shown in Fig.~\ref{fig:conlme-weights}f. Before the first iteration, no consistency constraint is active, and the solution reduces to pure Laplacian mesh editing. The retrieved nearest-neighbor configurations are therefore unreliable, as the current unit cell configurations have not yet been guided toward the feasible region of $\mathcal{D}$. As iterations proceed, the unit cell configurations progressively approach entries in $\mathcal{D}$, improving the quality of the retrieved nearest neighbors and allowing the consistency constraint to become increasingly effective. This, in turn, contracts the tails of the area distribution and reduces dissimilarity scores.

\section{Conditional diffusion model architecture, training, and performance}\label{app:cdm-model}
\subsection{Model architecture}
The diffusion process uses $T = 1000$ timesteps with a cosine noise schedule, and designs are generated by full ancestral sampling over all $T$ steps. The noisy input is first projected by a $7 \times 1$ stem convolution. The diffusion timestep is encoded via a sinusoidal embedding~\cite{vaswani_attention_2023} followed by a two-layer MLP with GELU activation, which modulates all ResNet~\cite{he_deep_2015} blocks through scale-shift conditioning. The encoder comprises four stages with channel progression $d \to 2d \to 4d \to 8d$~\cite{ronneberger_u-net_2015}, each applying cross-attention with $\boldsymbol{\theta}$, two time-conditioned ResNet blocks, linear attention, and downsampling. The bottleneck consists of cross-attention, a ResNet block, full self-attention with a single head, and a second ResNet block. The decoder mirrors the encoder with skip connections. The decoder output is fused with the stem residual, passed through a final ResNet block, and projected by a $1 \times 1$ convolution to produce the diffusion target. The hyperparameters are listed in Table~\ref{tab:cdm-hyperparameters}.

\subsection{Training}
The dataset of 59,999 samples is split into training and test sets at an 8:2 ratio. The Adam optimizer is used with a learning rate of $5\times10^{-5}$. Automatic mixed precision (AMP, fp16), gradient clipping (maximum norm = 1.0), and exponential moving average (EMA, decay = 0.995) are employed to stabilize optimization. The model is trained for 10,000 training steps with a batch size of 128 and two gradient accumulation steps (approximately 53 passes over the training set), and the full training process completes in less than one hour. The model is trained using Python 3.11.14 and PyTorch 2.8.0~\cite{paszke_pytorch_2019} with CUDA and cuDNN support. Hardware and resource allocations are summarized in Appendix~\ref{app:computational-setup}.

\begin{table}[pos=!htbp]
    \centering
    \caption{\textbf{Hyperparameters of the conditional diffusion model.}}
    \begin{tabular}{l|p{6cm}}
        \toprule
        Hyperparameter & Value \\
        \midrule
        ResNet blocks per downsampling stage & 2 \\
        ResNet blocks per upsampling stage & 2 \\
        Residual block normalization & RMS Normalization \\
        Activation function & SiLU (GELU in the time-embedding MLP) \\
        Time embedding & Sinusoidal positional embedding + 2-layer MLP \\
        Condition injection & Normalized $\boldsymbol{\theta}\in\mathbb{R}^{1\times 8}$ used directly as cross-attention context (no separate encoder) \\
        Attention head dimension & 32 \\
        Number of attention heads & 4 (1 in the bottleneck self-attention) \\
        Sequence resolutions & $8 \to 4 \to 2 \to 1$ \\
        Feature dimension & $d \to 2d \to 4d \to 8d$ \\
        Initial feature dimension & $d = 256$ \\
        Diffusion timesteps & $T = 1000$ \\
        Noise schedule & Cosine \\
        Prediction target & Noise $\boldsymbol{\epsilon}$ \\
        Sampling & Ancestral sampling over all $T$ steps \\
        \bottomrule
    \end{tabular}
    \label{tab:cdm-hyperparameters}
\end{table}
\subsection{Evaluation on the test set}
\begin{figure}[pos=!htbp]
    \centering
    \includegraphics[width=0.8\linewidth]{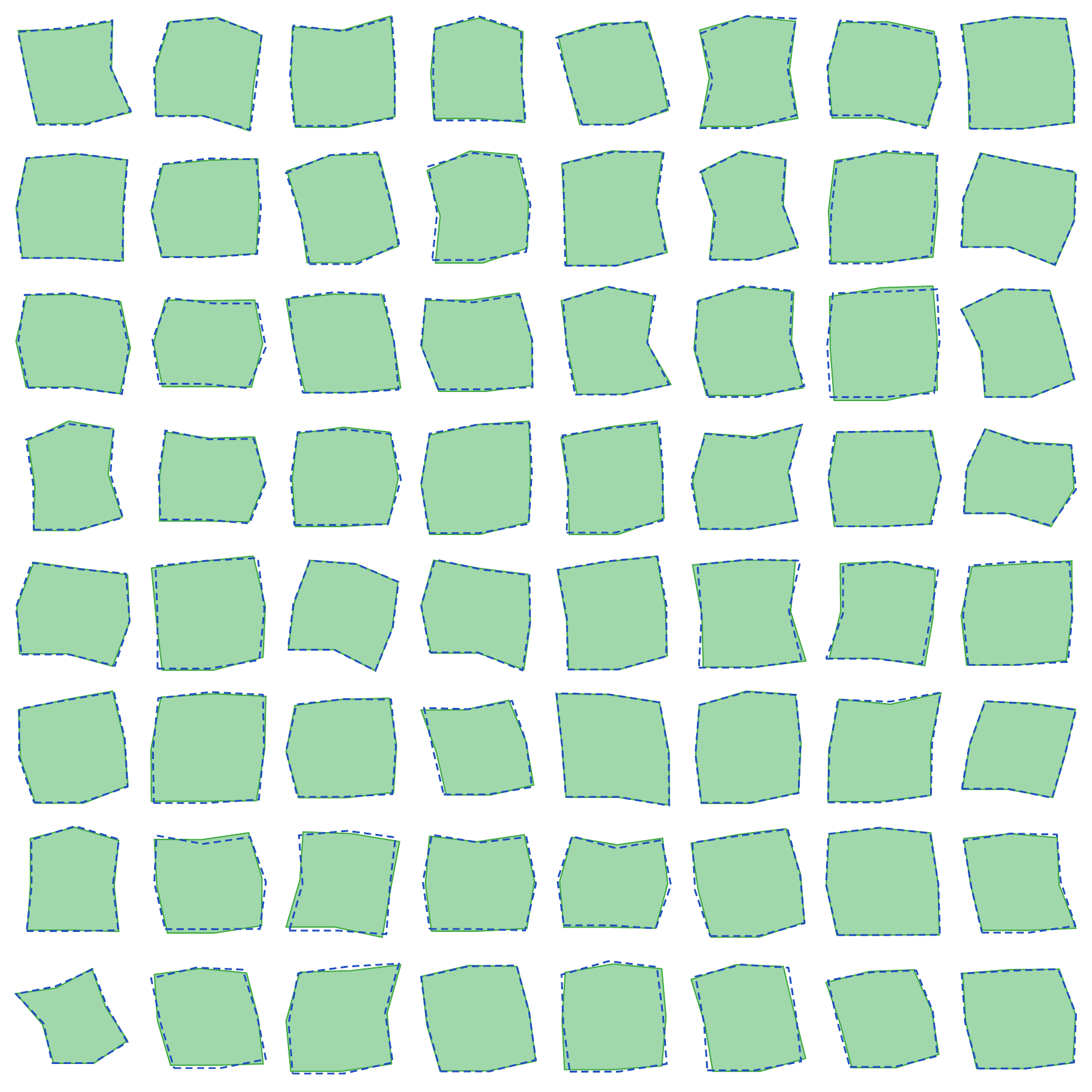}
    \caption{\textbf{Comparison of the target deformation configurations sampled from the test set and the simulated configurations of generated unit cells.} The blue dashed line presents the target configurations, while the filled green shapes present the simulated configurations.}
    \label{fig:cdm-performance}
\end{figure}

To evaluate the cDM, we randomly selected $K=1{,}000$ samples from the test set. For each sample, the target response configuration $\boldsymbol{\theta}$ and randomly sampled initial noise were fed into the model to generate candidate infill designs. The generated parameters were clipped to the predefined feasible design ranges to ensure geometric validity, and FEA (Appendix~\ref{app:fem}) was performed on the resulting structures to obtain their deformation responses; all 1,000 simulations converged. The accuracy is measured by the coefficient of determination between the target angle changes and the simulated angle changes, pooled over the eight angles of all samples:
\begin{equation}
    \begin{aligned}
        R^2_{\mathrm{test}} = 1-\frac{\sum_{i=1}^K\sum_{j=1}^8 (\Delta\theta_{ij}^*-\Delta\theta_{ij})^2}{\sum_{i=1}^K\sum_{j=1}^8 (\Delta\theta_{ij}^*-\Delta\bar{\theta}_{j}^*)^2},
    \end{aligned}
    \label{Eq:r2-test}
\end{equation}
where $\Delta \theta^*_{ij}$ is the target change of the $j$-th angle of the $i$-th sample, $\Delta\theta_{ij}$ is the corresponding simulated change, and $\Delta\bar{\theta}_{j}^*=\frac{1}{K}\sum_{i=1}^K\Delta\theta_{ij}^*$ is the mean target change of the $j$-th angle. Because the test samples span the whole dataset, the target angle changes vary widely, and the coefficient of determination is a meaningful measure of how much of this variation the model reproduces. The cDM achieves $R^2_{\mathrm{test}}=0.91$ on the test set, corresponding to $\mathrm{RMS}_{\mathrm{micro}}=3.41^\circ$ (Eq.~\eqref{Eq:rms-micro}). Fig.~\ref{fig:cdm-performance} visualizes 64 randomly selected samples, showing that the majority of generated designs closely match their target configurations, with only a small number exhibiting noticeable deviations. Overall, the model achieves sufficient accuracy for the microscale inverse design stage.

\section{Metrics of performance evaluation}\label{sec:metrics}
We use different metrics to evaluate the performance of the proposed framework at two scales. At the microscale, the root-mean-square (RMS) error of the angle changes, taken over the eight angles of all unit cells, is used to evaluate the batch of designs by
\begin{equation}
    \begin{aligned}
        \mathrm{RMS}_{\mathrm{micro}} = \sqrt{\frac{1}{8K}\sum_{i=1}^K\sum_{j=1}^8 (\Delta\theta_{ij}^*-\Delta\theta_{ij})^2},
    \end{aligned}
    \label{Eq:rms-micro}
\end{equation}
where $K$ is the number of unit cells, $\Delta \theta^*_{ij}$ is the target angle change at the $j$-th angle of the $i$-th unit cell, according to the local configurations solved in the macroscale subproblem, and $\Delta\theta_{ij}$ is the corresponding actual angle change achieved by the solutions of the microscale subproblem, obtained from FEA of the designed unit cell.
Unlike a coefficient of determination, $\mathrm{RMS}_\mathrm{micro}$ does not depend on the spread of the target angle changes, so designs with small and large local deformations are measured on the same absolute scale.
$\mathrm{RMS}_\mathrm{micro}$ measures how closely the designed unit cells, simulated in isolation, reproduce the configurations required by the macroscale solution.
A large $\mathrm{RMS}_\mathrm{micro}$ may indicate that the macroscale solution contains out-of-distribution configurations, e.g., too large a deformation or an unrealistic cell configuration. A small $\mathrm{RMS}_\mathrm{micro}$, however, does not by itself imply a small macroscale error (Section~\ref{sec:error-sources}), which is why the two scales are evaluated separately.

At the macroscale, the mean absolute error ($\mathrm{MAE}$) between the target displacement $\boldsymbol{d}^\mathrm{target}$ and the actual displacement $\boldsymbol{d}^\mathrm{actual}$ (obtained from finite element analysis) at the handle vertices is used to evaluate the global design performance:
\begin{equation}
    \begin{aligned}
        \mathrm{MAE} = \frac{1}{N_h}\sum_{i=1}^{N_h}\left \|\boldsymbol{d}^\mathrm{target}_i-\boldsymbol{d}^\mathrm{actual}_i\right\|,
    \end{aligned}
\end{equation}
where $N_h$ is the number of handle vertices. 

To more directly reflect the relative magnitude of the error, we take the maximum target displacement in each example as the characteristic length, $L_{\mathrm{c}}$, and define the mean relative error (MRE) as
\begin{equation}
    \mathrm{MRE}=\frac{\mathrm{MAE}}{L_{\mathrm{c}}}.
\end{equation}
As a supplementary metric, the $R^2$ score between the target positions and the actual positions of the handle vertices is also used to assess design performance. The $R^2$ score is computed independently for the two coordinate dimensions and then averaged to obtain the final metric:
\begin{equation}
    \begin{aligned}
        R^2_\mathrm{macro} = \frac{1}{2}\sum_{j=1}^2\left (1-\frac{\sum_{i=1}^{N_h} (v_{ij}^\mathrm{actual}-v_{ij}^\mathrm{target})^2}{\sum_{i=1}^{N_h} (v_{ij}^\mathrm{target}-\bar{v}_{j}^\mathrm{target})^2}\right ),
    \end{aligned}
\end{equation}
where $v^\mathrm{target}_{ij}$ is the target position for the $j$-th coordinate of handle vertex $i$, $v^\mathrm{actual}_{ij}$ is the corresponding actual handle vertex position obtained from finite element analysis, and $\bar{v}_{j}^\mathrm{target}$ is the mean of $v^\mathrm{target}_{ij}$ over all handle vertices.

\section{Computational setup}\label{app:computational-setup}
The diffusion model was trained on a Dell Precision 7960 Tower Workstation using a single NVIDIA RTX A4000 GPU (16 GB VRAM), four Intel Xeon w5-3423 CPU cores, and 32 GB of RAM. The sinusoidal beam benchmark (Section~\ref{sec:scalability}), including the wall-clock times reported in Figs.~\ref{fig:scalability}j and m and the full-structure FEA, was run on the same workstation using a single RTX A4000 GPU and the 12-core Intel Xeon w5-3423 CPU, with one design configuration computed at a time. All other SAM design computations and full-structure FEA were performed on a separate system using a single NVIDIA A100 GPU, four Intel Xeon 6248R CPU cores, and 128 GB of RAM.

\section{Time complexity analysis}\label{app:time-complexity}
To compare the computational efficiency of the SAM framework with FEM-based optimization methods, we conduct a time-complexity analysis for the design of a two-dimensional metamaterial comprising an $N \times N$ array of unit cells, i.e., $N_c = N^2$ unit cells in total. Topology optimization (TO) is selected as the representative baseline, as it is generally more computationally efficient than alternative optimization approaches such as genetic algorithms.

For TO, the number of iterations required for convergence is problem-dependent; we denote this as $k$. The computational domain is discretized using quadrilateral finite elements, with each unit cell meshed using an $m \times m$ finite element grid.

For the weakly coupled thermomechanical problem considered here, the dominant computational cost at each iteration arises from solving the finite element system. Employing a banded direct solver based on Gaussian elimination, the complexity of each solve scales as $O(nw^2)$~\cite{farmaga_evaluation_2011}, where $n$ denotes the number of degrees of freedom and $w$ is the bandwidth of the global stiffness matrix. For the uniform grid considered here, with $(mN+1)^2$ nodes and two displacement degrees of freedom per node, $n = 2(mN+1)^2$, and the bandwidth under row-wise node numbering is $w \approx 2(mN+1)$. 

The overall computational complexity of TO over $k$ iterations, therefore, scales as
\begin{equation}
    O\!\left(k\,n\,w^2\right) = O\!\left(8k(mN + 1)^4\right)\approx O(km^4N^4) = O(km^4N_c^2),
\end{equation}
which grows quadratically with the total number of unit cells $N_c$. 

The banded direct solver is one representative choice and is not a lower bound on the cost of TO. With a sparse direct solver based on nested dissection, a two-dimensional solve costs $O(n^{1.5})$, giving $O(km^3N_c^{1.5})$; with an optimal iterative solver such as multigrid, a solve costs $O(n)$, giving $O(km^2N_c)$. In the latter case TO also scales linearly with $N_c$, and it differs from SAM by the factor $km^2$ associated with the repeated finite element analyses of the assembled structure.

For the SAM framework, no finite element simulation is required, and the computational complexity decomposes into two stages. In the macroscale solving stage, the time complexity of Laplacian mesh editing scales linearly with the number of vertices \cite{mm_mesh_2016}, i.e.,
\begin{equation}
    O(N^2) = O(N_c),
\end{equation}
since the number of vertices in the mesh is proportional to $N^2$.
Two further costs belong to this stage. First, each ConLME iteration retrieves the nearest configuration of every unit cell by exhaustive search over the dataset, which costs $O(N_c|\mathcal{D}|)$; since the dataset size $|\mathcal{D}|$ is fixed, this is linear in $N_c$. Second, the number of ConLME iterations is capped at a prescribed maximum that does not depend on the resolution, so it enters as a constant factor. The linear cost of the sparse solve is not guaranteed in general (a sparse direct factorization of a two-dimensional mesh costs $O(N_c^{1.5})$); for the resolutions considered here, up to 1328 unit cells, the measured time of the macroscale stage grows linearly (Fig.~\ref{fig:scalability}j).

In the microscale design stage, the cDM-based approach is taken as a representative example, generating one infill design for each of the $N_c$ unit cells. Assuming each batch accommodates at most $C$ samples and that generating a single sample incurs a complexity of $O(T\,C_{\text{model}})$, where $T$ denotes the number of diffusion steps and $C_{\text{model}}$ represents the cost of a single U-Net forward pass, the overall complexity of the microscale design stage is given by
\begin{equation}
    O\!\left(\left\lceil \frac{N_c}{C} \right\rceil \, T \, C_{\text{model}} \right).
\end{equation}
Under fixed batch size and diffusion settings, the overall time complexity combining both stages is
\begin{equation}
    O(N^2)+O\!\left(\left\lceil \frac{N_c}{C} \right\rceil \, T \, C_{\text{model}} \right)\approx O(N_c).
\end{equation}

This approximation holds because the number of diffusion steps $T$ and the U-Net cost $C_{\text{model}}$ are fixed constants independent of the structural resolution. The design time of the cDM-based SAM therefore scales linearly with the total number of unit cells, $O(N_c)$, compared with $O(km^4N_c^2)$, $O(km^3N_c^{1.5})$, or $O(km^2N_c)$ for TO, depending on the solver. The linear scaling of SAM is supported by the measured runtimes in the tested range rather than by an asymptotic guarantee. The analysis covers the online design cost only: the dataset of 60{,}000 nonlinear unit cell simulations (about 0.6~s each, including meshing) and the training of the diffusion model (less than one hour) are one-time offline costs, which are reused across design domains and are not included.
\section{Comparison of domain discretization methods}\label{app:discretization}
The decomposition strategy adopted in this work treats a closed-loop linkage composed of multiple bars as the minimal design unit. Compared with the conventional approach that discretizes the design domain into bar-based structures connected by ideal hinges \cite{walker_computational_2024, liwei_2025}, we demonstrate, using a simple example, why SAM uses unit-level discretization.

\subsection{Problem setup}
\begin{figure}[pos=!htbp]
    \centering
    \includegraphics[width=0.5\linewidth]{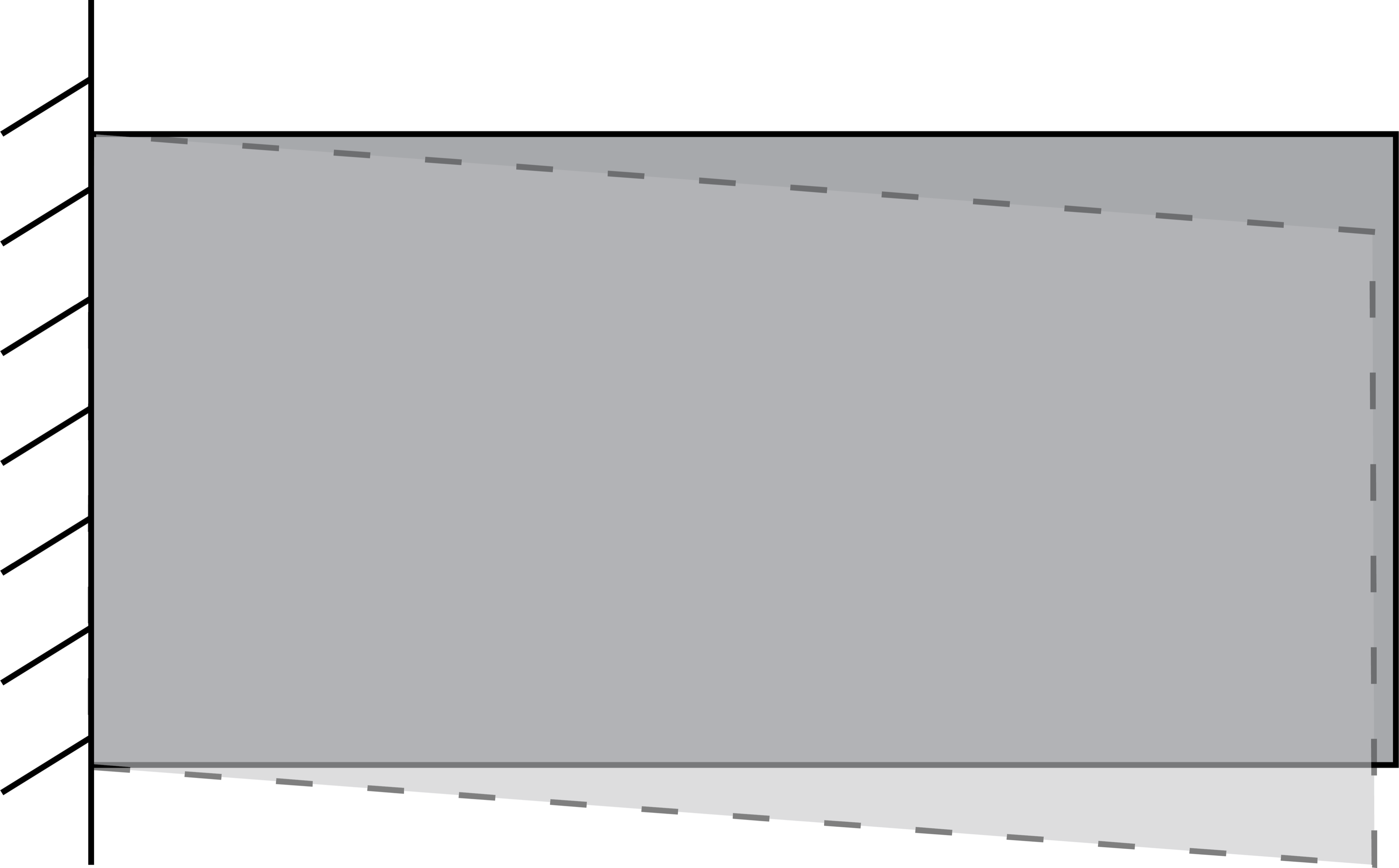}
    \caption{\textbf{A simple shape-morphing problem.}}
    \label{fig:bar-problem}
\end{figure}

Given a rectangular material domain of length 2 mm and width 1 mm as shown in Fig.~\ref{fig:bar-problem}, with its left boundary fixed, we design the embedded active material using either bars or unit cells as the basic design elements, such that the upper-right corner undergoes a downward displacement of $d=0.2$ mm. We consider a simple solution where this domain is discretized into two equal squares.

The solution of a bar-based design can be manually derived using nine bars, as shown in Fig.~\ref{fig:bar-vs-cell}, where the gray outer bars are passive members, and the blue bars are active members made of the same material. Under this setup, the initial length of each passive bar is $L_p = 1\,\mathrm{mm}$, while the initial length of each active bar is $L_a = 1.414\,\mathrm{mm}$. We can solve this simple bar-based design problem analytically. Assuming that the passive material does not deform and that all active members undergo the same strain, the deformed angle $\beta$ is given by
\begin{equation}
    \beta^{*}=\frac{\pi}{2}+\arcsin\!\left(\frac{d}{2}\right)=1.670963.
\end{equation}
Accordingly, the deformed length of the blue bars becomes
\begin{equation}
    L_a^{*}=2L_p\sin\!\left(\frac{\beta^{*}}{2}\right)=1.4832.
\end{equation}
The resulting strain in each active bar is therefore
\begin{equation}
    \epsilon=\frac{L_a^{*}-L_a}{L_a}=0.0488.
\end{equation}
To realize this target strain through thermal expansion, we set the coefficient of thermal expansion of the active material to $4.88\times10^{-4}\,^\circ\mathrm{C}^{-1}$ and apply a temperature increase of $100^\circ\mathrm{C}$, followed by a finite element simulation.

For the unit-cell-based discretization, we adopt the AS-based SAM setup described in Section~\ref{sec:as} for inverse design.

\subsection{Results of SAM and bar-based framework}
As shown in Fig.~\ref{fig:bar-vs-cell}, finite element simulations indicate that the deformation error of the bar-based design reaches 31.4\% compared to the theoretically achievable configuration, whereas the error of the AS-based SAM design is 7.7\%, which is markedly smaller. This discrepancy arises because the bar-based design neglects the interactions between adjacent bars and considers only axial deformation. In practice, this simplification introduces significant errors. Moreover, as the number of bars increases, the likelihood of failure also grows, as evidenced by the buckling observed in some of the structures presented in \cite{liwei_2025}.
In contrast, the unit-cell-based design implicitly incorporates the interactions among structural elements through the dataset of unit-cell configurations. As a result, when assembled, it is able to achieve more accurate and stable deformation behavior.

\subsection{Deformation mode ambiguity in bar-based design frameworks}
\begin{figure}[pos=!htbp]
    \centering
    \includegraphics[width=1.0\linewidth]{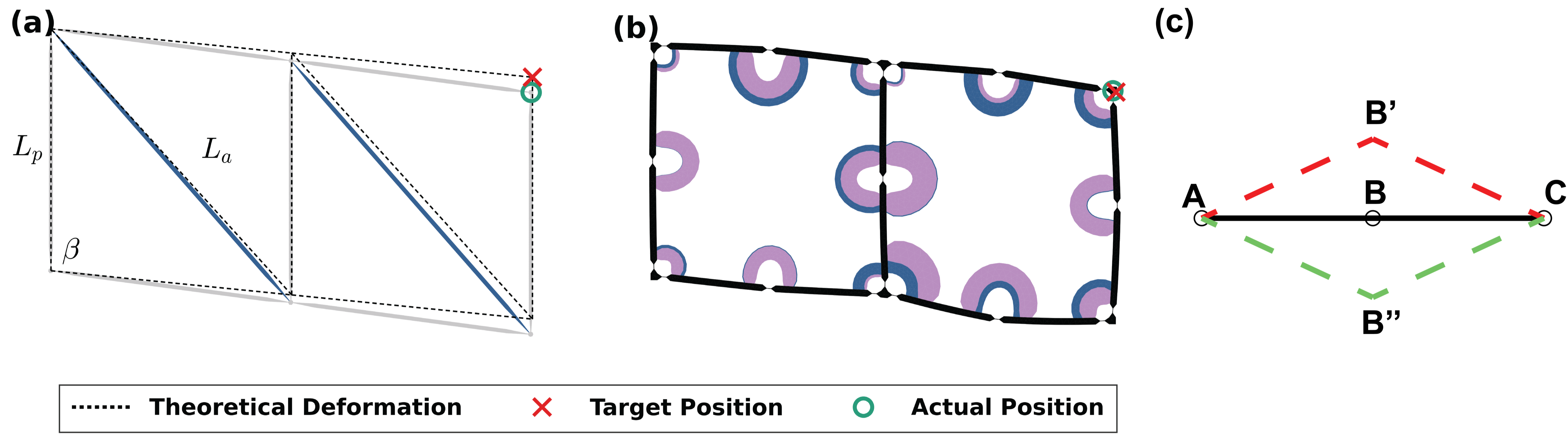}
    \caption{\textbf{Comparison of bar-based and cell-based discretizations.} 
    \textbf{a,} The simulation results of bar-based element design.
    \textbf{b,} The simulation results of unit-cell-based element design (using AS-based SAM).
    \textbf{c,} The demonstration of deformation mode ambiguity in bar-based design frameworks.
    }
    \label{fig:bar-vs-cell}
\end{figure}

Another fundamental limitation of bar-based design frameworks is the inherent ambiguity in their deformation modes. Consider the two-bar system shown in Fig.~\ref{fig:bar-vs-cell}c, where both ends A and C are fixed, and the design intent is to displace the intermediate node B upward (red path). However, the same bar design is equally compatible with a downward displacement of B (green path), since axial deformation alone carries no directional information. As a result, the realized deformation mode depends on the assembly configuration rather than the design itself, introducing an irresolvable ambiguity. The unit-cell-based approach adopted in the SAM framework eliminates this ambiguity by assigning each cell a unique, directionally encoded deformation mode, ensuring that the intended deformation is unambiguously realized upon assembly.

\section{More examples}\label{app:examples}
\begin{figure}[pos=!htbp]
    \centering
    \includegraphics[width=1.0\linewidth]{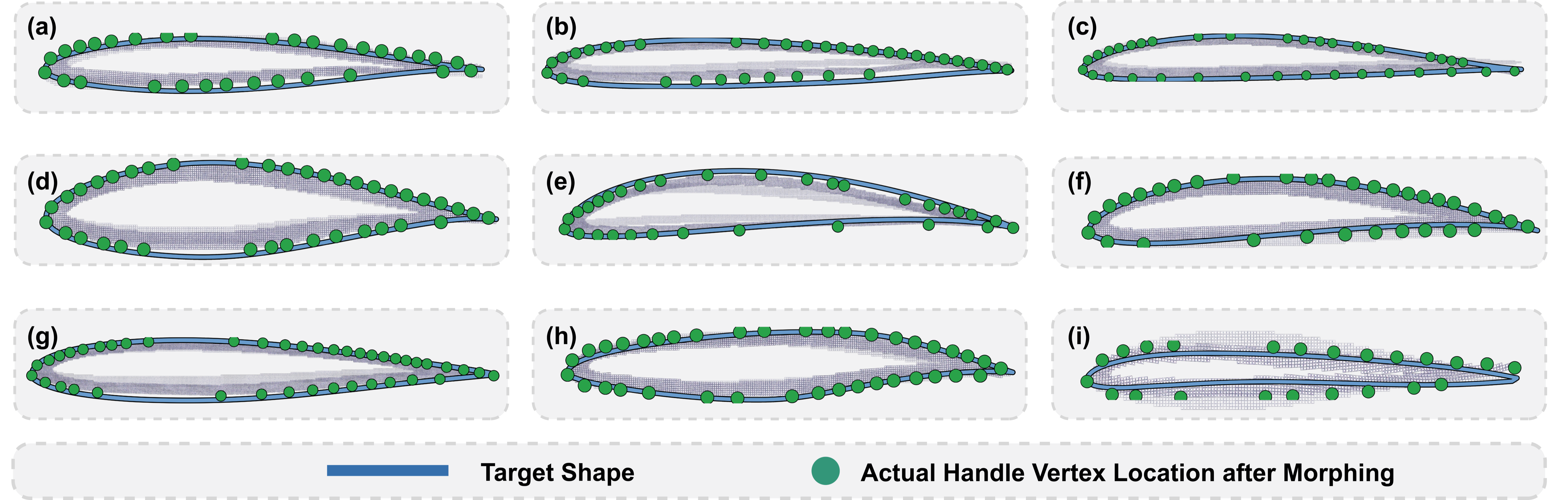}
    \caption{\textbf{More airfoil shape matching examples.} The green dots indicate the deformed positions, and the blue curves indicate the target positions.}
    \label{fig:airfoil-examples}
\end{figure}

This section presents additional shape-morphing examples to further demonstrate the generality of the SAM framework. Nine pairs of initial and target airfoil shapes were selected from the dataset~\cite{chen_airfoil_2020, selig_uiuc_1996} and designed using the cDM-based SAM framework, with FEA results shown in Fig.~\ref{fig:airfoil-examples} and quantitative metrics summarized in Table~\ref{tab:airfoil-examples}. The mean relative error lies between 14\% and 28\% for the nine cases, with a mean of 22\%, comparable to the airfoil examples of Table~\ref{tab:performance}. All nine airfoils are held in the same way: at one leading-edge vertex and at the handle vertex farthest from it among those whose prescribed displacement is below one unit-cell length. As discussed in Section~\ref{sec:generality}, the prescribed displacements are small compared with the chord, so that a support confined to the leading edge allows a small rigid rotation of the whole airfoil to dominate the error; the long baseline between the two fixed vertices suppresses this rotation. The same fixed vertices are used in the macroscale solution and in the FEA. Each case is a single design of the cDM-based framework.

\begin{table}[pos=!htbp]
\centering
\caption{Quantitative error statistics of all examples.}
\label{tab:airfoil-examples}
\begin{tabular}{l c c c c c c c}
\toprule
Case & $N_c$ & $N_h$& $\mathrm{MAE}$ (mm) & $R^2_\mathrm{macro}$ & $L_{\mathrm{c}}$ (mm) & $\mathrm{MRE}$\\
\midrule
Fig.~\ref{fig:airfoil-examples}a& 2524 & 31& 2.2116 & 0.9735 & 7.99 & 27.68\%\\
Fig.~\ref{fig:airfoil-examples}b& 3408 & 36& 1.6752 & 0.9797 & 7.50 & 22.33\%\\
Fig.~\ref{fig:airfoil-examples}c& 4104 & 36& 0.9472 & 0.9967 & 5.08 & 18.64\%\\
Fig.~\ref{fig:airfoil-examples}d& 3588 & 38& 1.1825 & 0.9969 & 6.05 & 19.56\%\\
Fig.~\ref{fig:airfoil-examples}e& 4267 & 31& 2.4699 & 0.9892 & 13.51 & 18.28\%\\
Fig.~\ref{fig:airfoil-examples}f& 3294 & 33& 1.4309 & 0.9886 & 7.02 & 20.38\%\\
Fig.~\ref{fig:airfoil-examples}g& 4484 & 42& 1.3973 & 0.9942 & 9.88 & 14.14\%\\
Fig.~\ref{fig:airfoil-examples}h& 3215 & 38& 1.2634 & 0.9916 & 4.83 & 26.18\%\\
Fig.~\ref{fig:airfoil-examples}i& 1354 & 23& 2.6095 & 0.7636 & 9.16 & 28.49\%\\
\bottomrule
\end{tabular}
\end{table}
The largest errors occur in cases a, h, and i. Case i is the smallest airfoil of the set, discretized with the fewest unit cells, and its $R^2_\mathrm{macro}$ is the lowest of the nine cases.

\section{Proof of volume reduction}\label{app:volume}
\begin{figure}[pos=!htbp]
    \centering
    \includegraphics[width=0.8\linewidth]{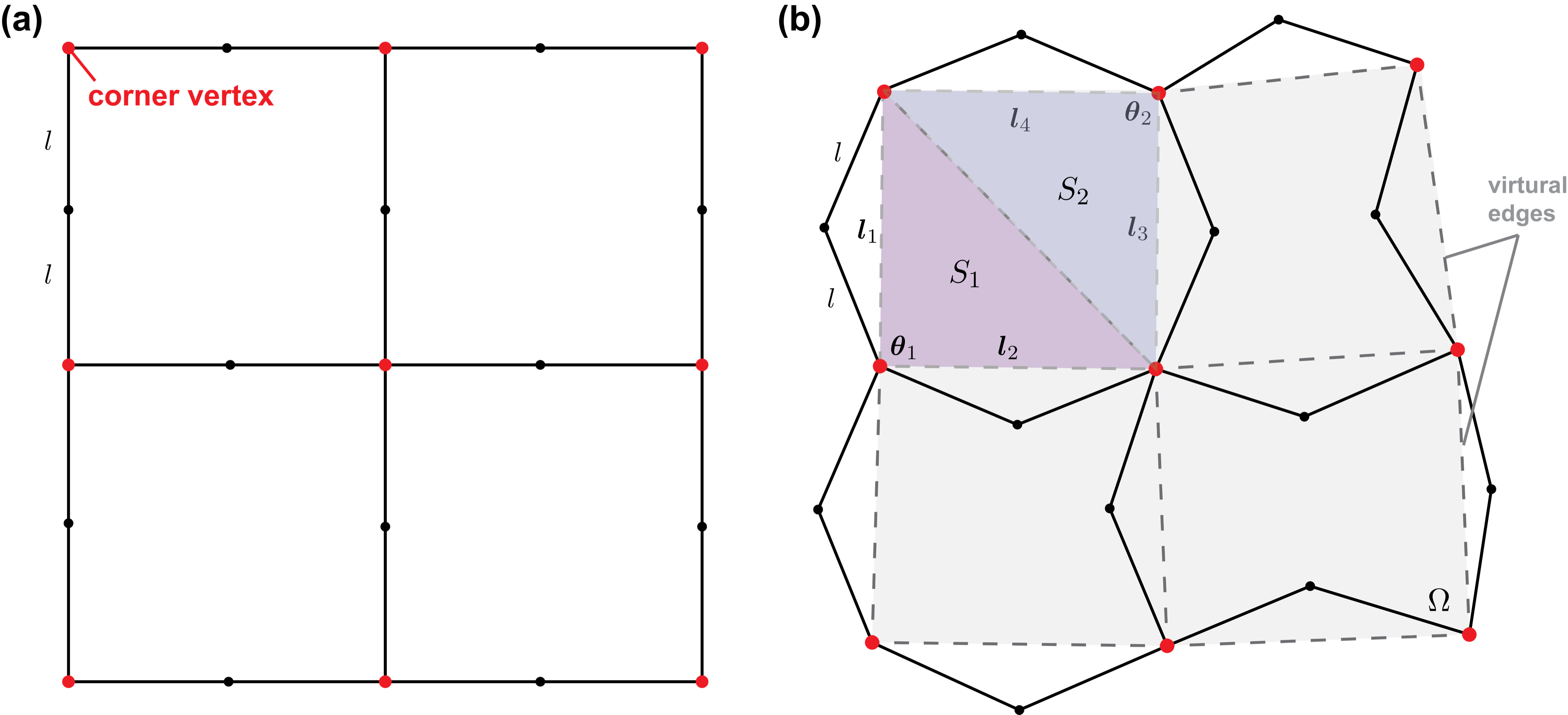}
    \caption{\textbf{Proof of compactness.}
     \textbf{a,} The initial mesh.
     \textbf{b,} The deformed mesh.}
    \label{fig:volume-proof}
\end{figure}

We provide the proof of the volume reduction properties of SAM. The result concerns an idealized setting: rigid bars, densely arranged cells, and the region $\Omega$ formed by the quadrilaterals spanned by the four corner vertices of each unit cell. It does not cover the hinge vertices on the outer boundary, which may bulge outward, and therefore does not bound the outer envelope of a finite structure or the space swept during actuation. To obtain the theoretical area of the deformed metamaterial, we consider an ideal case with just the undetermined mechanism, as shown in Fig.~\ref{fig:volume-proof}a. Each bar has a length of $l$. So the initial area for $n$ square cells is,
\begin{equation}
    \begin{aligned}
        S_0 = 4nl^2.
    \end{aligned}
\end{equation}
Once the mechanism deformed as shown in Fig.~\ref{fig:volume-proof}b, the shape of the unit cell will become irregular. By connecting the corner vertices of each cell in sequence, a set of virtual boundaries is formed, which enclose the shaded region $\Omega$ shown in Fig.~\ref{fig:volume-proof}b. Apparently, the length of an arbitrary virtual edge of one unit cell should be smaller than the length of a square in the original configuration, e.g., $l_i<2l$. Since the boundaries are negligible when the number of unit cells increases, we could just consider the area of region $\Omega$. In this case, the area of the deformed region can be computed by summing the areas of individual cells, each defined by the regions enclosed by their corner points. Thus, the total area of the arbitrary region could be calculated by,
\begin{equation}
    \begin{aligned}
        \widetilde{S} &= \sum_{i=1}^n (S_1^i+S_2^i)=\sum_{i=1}^n\left( \frac{1}{2}l_1^il_2^i\sin\theta_1^i+\frac{1}{2}l_3^il_4^i\sin\theta_2^i\right)\\
        &\leq \sum_{i=1}^n \frac{1}{2}4l^2\left(\sin\theta_1^i+\sin\theta_2^i\right) \\
        &\leq \sum_{i=1}^n 4l^2 =4nl^2=S_0.
    \end{aligned}
\end{equation}
Thus, the total area only reduces after deformation.
\section{Unit cell deformation in the assembled structure}\label{app:assembly}
This appendix gives the cell-by-cell comparison that underlies the discussion of the assembly effect in Section~\ref{sec:error-sources}. It considers the sinusoidal beam benchmark at the highest resolution, $N_c = 1328$ unit cells, and the largest amplitude, $A = 1.5$, for which the macroscale error of AS-based SAM is higher than that of cDM-based SAM although its microscale error is lower.

The beam deflects mainly by shearing its unit cells. The left and right edges of a cell remain nearly parallel to their initial direction, while its bottom and top edges rotate. We therefore measure the shear $\gamma$ of a unit cell as the mean rotation of its bottom and top edges minus the mean rotation of its left and right edges, computed from the positions of its four corner vertices. This quantity is unaffected by a rigid rotation of the cell and equals the local slope that the cell contributes to the beam.

For every unit cell, three values are compared. The first is the shear $\gamma^{*}$ required by the macroscale solution. The second is the shear $\gamma_{\mathrm{iso}}$ delivered by the designed unit cell when it is simulated in isolation, as in the dataset. The third is the shear $\gamma_{\mathrm{asm}}$ of the same cell in the finite element solution of the assembled structure, obtained from the displacements of its corner vertices. In these terms, the microscale error compares $\gamma_{\mathrm{iso}}$ with $\gamma^{*}$, the assembly effect is the change from $\gamma_{\mathrm{iso}}$ to $\gamma_{\mathrm{asm}}$, and the macroscale error results from the difference between $\gamma_{\mathrm{asm}}$ and $\gamma^{*}$ accumulated over the structure.

\begin{figure}[hbt!]
    \centering
    \includegraphics[width=\linewidth]{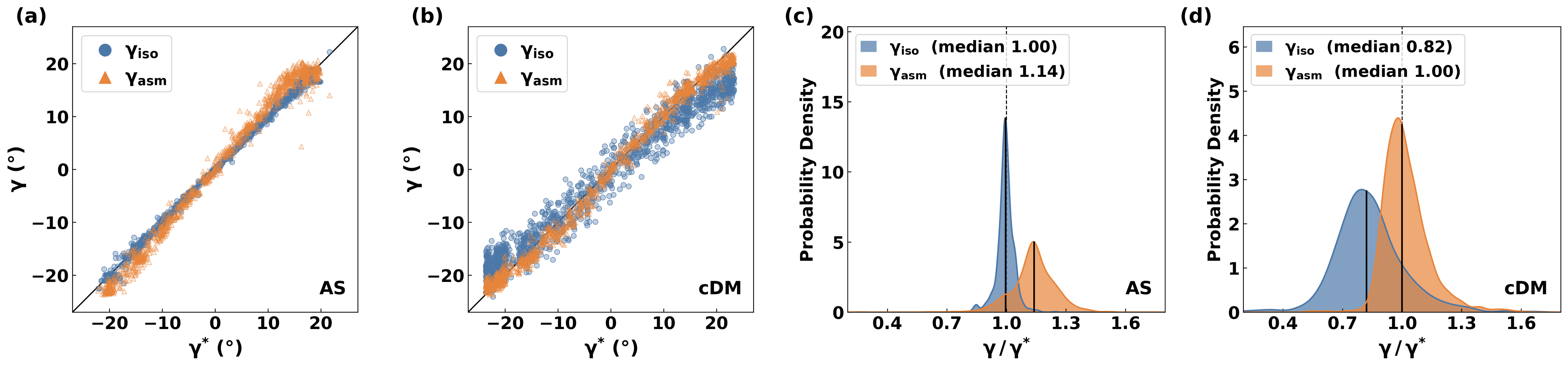}
    \caption{\textbf{Shear of the unit cells of the sinusoidal beam in isolation and inside the assembled structure} ($N_c = 1328$, amplitude $A = 1.5$). $\gamma^{*}$ is the shear required by the macroscale solution, $\gamma_{\mathrm{iso}}$ the shear of the designed unit cell simulated in isolation, and $\gamma_{\mathrm{asm}}$ the shear of the same cell in the assembled structure.
    \textbf{a, b,} $\gamma_{\mathrm{iso}}$ and $\gamma_{\mathrm{asm}}$ against $\gamma^{*}$ for every unit cell of the AS-based (\textbf{a}) and of the cDM-based (\textbf{b}) design. The line marks equality.
    \textbf{c, d,} Distributions of $\gamma_{\mathrm{iso}}/\gamma^{*}$ and $\gamma_{\mathrm{asm}}/\gamma^{*}$ for the AS-based (\textbf{c}) and the cDM-based (\textbf{d}) design, over the cells with $|\gamma^{*}|$ above 30\% of its largest value. Solid vertical lines mark the medians, and the dashed line marks a ratio of one. One of the three cDM-based designs is shown.}
    \label{fig:assembly}
\end{figure}

Fig.~\ref{fig:assembly} shows the result. In isolation, the unit cells retrieved by AS deliver the required shear, with a median of $\gamma_{\mathrm{iso}}/\gamma^{*}$ of 1.00 (Figs.~\ref{fig:assembly}a and \ref{fig:assembly}c). Those generated by the cDM deliver less, with a median of 0.82 and a wider distribution (Figs.~\ref{fig:assembly}b and \ref{fig:assembly}d).

This underestimation is not specific to the beam. On the test set of Appendix~\ref{app:cdm-model}, the designs generated by the cDM deliver on average 0.87 of the required shear. The underestimation is consistent with the distribution of the training data, which is sparse at extremely large deformations. The shear of the unit cells in the dataset exceeds $20^\circ$ in only 0.03\% of the samples. The shear required at the largest amplitude reaches $23^\circ$, the largest value in the dataset, which is why the underestimation is most pronounced there.

Inside the assembled structure, the unit cells of both methods shear more than in isolation. We attribute this increase to the sharing of the frame. In isolation, the active infill of a unit cell works against the stiffness of its own frame members and hinges, whereas in the assembled structure each interior member and hinge is shared with a neighboring unit cell and is driven by the infills of both. For the AS-based design, the median of $\gamma_{\mathrm{asm}}/\gamma^{*}$ rises to 1.14, so that the unit cells exceed the required shear by about 14\% (Fig.~\ref{fig:assembly}c). For the cDM-based design, the median rises from 0.82 to 1.00 (Fig.~\ref{fig:assembly}d). The underestimation of the isolated unit cells thus largely offsets the increase in the assembly, which is why the cDM-based design reaches the lower macroscale error in this configuration.

\bibliographystyle{cas-model2-names}
\bibliography{references}

@article{cui_novel_2023,
    title = {A novel auxetic unit cell for {3D} metamaterials of designated negative {Poisson}'s ratio},
    volume = {260},
    issn = {0020-7403},
    url = {https://www.sciencedirect.com/science/article/pii/S0020740323005167},
    doi = {10.1016/j.ijmecsci.2023.108614},
    urldate = {2023-07-29},
    journal = {International Journal of Mechanical Sciences},
    author = {Cui, Jipeng and Zhang, Liangchi and Gain, Asit Kumar},
    year = {2023},
    pages = {108614},
}

@article{wang_mechanical_2022,
    title = {Mechanical cloak via data-driven aperiodic metamaterial design},
    volume = {119},
    url = {https://www.pnas.org/doi/abs/10.1073/pnas.2122185119},
    doi = {10.1073/pnas.2122185119},
    language = {EN},
    number = {13},
    urldate = {2024-04-02},
    journal = {Proceedings of the National Academy of Sciences},
    author = {Wang, Liwei and Boddapati, Jagannadh and Liu, Ke and Zhu, Ping and Daraio, Chiara and Chen, Wei},
    year = {2022},
    pages = {e2122185119},
}

@article{wenz_designing_2021,
    title = {Designing {Shape} {Morphing} {Behavior} through {Local} {Programming} of {Mechanical} {Metamaterials}},
    volume = {33},
    issn = {1521-4095},
    url = {https://onlinelibrary.wiley.com/doi/abs/10.1002/adma.202008617},
    doi = {10.1002/adma.202008617},
    number = {37},
    urldate = {2025-11-26},
    journal = {Advanced Materials},
    author = {Wenz, Franziska and Schmidt, Ingo and Leichner, Alexander and Lichti, Tobias and Baumann, Sascha and Andrae, Heiko and Eberl, Christoph},
    year = {2021},
    pages = {2008617},
}

@article{skarsetz_programmable_2022,
    title = {Programmable {Auxeticity} in {Hydrogel} {Metamaterials} via {Shape}-{Morphing} {Unit} {Cells}},
    volume = {9},
    copyright = {© 2022 The Authors. Advanced Science published by Wiley-VCH GmbH},
    issn = {2198-3844},
    url = {https://onlinelibrary.wiley.com/doi/abs/10.1002/advs.202201867},
    doi = {10.1002/advs.202201867},
    number = {23},
    urldate = {2025-11-26},
    journal = {Advanced Science},
    author = {Skarsetz, Oliver and Slesarenko, Viacheslav and Walther, Andreas},
    year = {2022},
    pages = {2201867},
}

@article{greenwood_soft_2025,
    title = {Soft multistable magnetic-responsive metamaterials},
    volume = {11},
    url = {https://www.science.org/doi/10.1126/sciadv.adu3749},
    doi = {10.1126/sciadv.adu3749},
    number = {29},
    urldate = {2025-11-26},
    journal = {Science Advances},
    author = {Greenwood, Taylor E. and Elder, Brian and Hasan, Md. Nahid and Anklam, Jared and Lee, Saebom and Teng, Jian and Wang, Pai and Kong, Yong Lin},
    year = {2025},
    pages = {eadu3749},
}

@article{zhao_encoding_2023,
    title = {Encoding reprogrammable properties into magneto-mechanical materials via topology optimization},
    volume = {9},
    number = {1},
    copyright = {2023 The Author(s)},
    issn = {2057-3960},
    url = {https://www.nature.com/articles/s41524-023-00980-2},
    doi = {10.1038/s41524-023-00980-2},
    urldate = {2025-11-26},
    journal = {npj Computational Materials},
    author = {Zhao, Zhi and Zhang, Xiaojia Shelly},
    year = {2023},
    pages = {57},
}

@article{lee_data-driven_2024,
    title = {Data-{Driven} {Design} for {Metamaterials} and {Multiscale} {Systems}: {A} {Review}},
    volume = {36},
    issn = {0935-9648, 1521-4095},
    url = {http://arxiv.org/abs/2307.05506},
    doi = {10.1002/adma.202305254},
    urldate = {2024-04-12},
    number = {8},
    journal = {Advanced Materials},
    author = {Lee, Doksoo and Chen, Wei Wayne and Wang, Liwei and Chan, Yu-Chin and Chen, Wei},
    year = {2024},
    pages = {2305254},
}

@article{jung_aperiodicity_2024,
    title = {Aperiodicity is all you need: {Aperiodic} monotiles for high-performance composites},
    volume = {73},
    issn = {1369-7021},
    url = {https://www.sciencedirect.com/science/article/pii/S1369702123004157},
    doi = {10.1016/j.mattod.2023.12.015},
    urldate = {2024-09-16},
    journal = {Materials Today},
    author = {Jung, Jiyoung and Chen, Ailin and Gu, Grace X.},
    year = {2024},
    pages = {1--8},
}

@article{wang_ih-gan_2022,
    title = {{IH}-{GAN}: {A} conditional generative model for implicit surface-based inverse design of cellular structures},
    volume = {396},
    issn = {0045-7825},
    shorttitle = {{IH}-{GAN}},
    url = {https://www.sciencedirect.com/science/article/pii/S0045782522002699},
    doi = {10.1016/j.cma.2022.115060},
    urldate = {2024-09-29},
    journal = {Computer Methods in Applied Mechanics and Engineering},
    author = {Wang, Jun and Chen, Wei (Wayne) and Da, Daicong and Fuge, Mark and Rai, Rahul},
    year = {2022},
    pages = {115060},
}

@article{wang_deep_2020,
    title = {Deep generative modeling for mechanistic-based learning and design of metamaterial systems},
    volume = {372},
    issn = {0045-7825},
    url = {https://www.sciencedirect.com/science/article/pii/S0045782520305624},
    doi = {10.1016/j.cma.2020.113377},
    urldate = {2025-12-17},
    journal = {Computer Methods in Applied Mechanics and Engineering},
    author = {Wang, Liwei and Chan, Yu-Chin and Ahmed, Faez and Liu, Zhao and Zhu, Ping and Chen, Wei},
    year = {2020},
    pages = {113377},
}

@article{choi_programming_2019,
    title = {Programming shape using kirigami tessellations},
    volume = {18},
    number = {9},
    copyright = {2019 The Author(s), under exclusive licence to Springer Nature Limited},
    issn = {1476-4660},
    url = {https://www.nature.com/articles/s41563-019-0452-y},
    doi = {10.1038/s41563-019-0452-y},
    urldate = {2025-03-09},
    journal = {Nature Materials},
    author = {Choi, Gary P. T. and Dudte, Levi H. and Mahadevan, L.},
    year = {2019},
    pages = {999--1004},
}

@article{wang_physics-aware_2023,
    title = {Physics-aware differentiable design of magnetically actuated kirigami for shape morphing},
    volume = {14},
    number = {1},
    copyright = {2023 The Author(s)},
    issn = {2041-1723},
    url = {https://www.nature.com/articles/s41467-023-44303-x},
    doi = {10.1038/s41467-023-44303-x},
    urldate = {2024-12-02},
    journal = {Nature Communications},
    author = {Wang, Liwei and Chang, Yilong and Wu, Shuai and Zhao, Ruike Renee and Chen, Wei},
    year = {2023},
    pages = {8516},
}

@article{li_geometric_2024,
    title = {Geometric mechanics of kiri-origami-based bifurcated mechanical metamaterials},
    volume = {382},
    issn = {1364-503X},
    url = {https://doi.org/10.1098/rsta.2024.0010},
    doi = {10.1098/rsta.2024.0010},
    number = {2283},
    urldate = {2025-12-17},
    journal = {Philosophical Transactions of the Royal Society A: Mathematical, Physical and Engineering Sciences},
    author = {Li, Yanbin and Zhou, Caizhi and Yin, Jie},
    year = {2024},
    pages = {20240010},
}

@article{tang_programmable_2017,
    title = {Programmable {Kiri}-{Kirigami} {Metamaterials}},
    volume = {29},
    number = {10},
    issn = {1521-4095},
    url = {https://onlinelibrary.wiley.com/doi/abs/10.1002/adma.201604262},
    doi = {10.1002/adma.201604262},
    urldate = {2025-10-05},
    journal = {Advanced Materials},
    author = {Tang, Yichao and Lin, Gaojian and Yang, Shu and Yi, Yun Kyu and Kamien, Randall D. and Yin, Jie},
    year = {2017},
    pages = {1604262},
}

@article{imediegwu_mechanical_2023,
    title = {Mechanical characterisation of novel aperiodic lattice structures},
    volume = {229},
    issn = {0264-1275},
    url = {https://www.sciencedirect.com/science/article/pii/S0264127523003374},
    doi = {10.1016/j.matdes.2023.111922},
    urldate = {2025-12-17},
    journal = {Materials \& Design},
    author = {Imediegwu, Chikwesiri and Clarke, Daniel and Carter, Francesca and Grimm, Uwe and Jowers, Iestyn and Moat, Richard},
    year = {2023},
    pages = {111922},
}

@article{liu_growth_2022,
    title = {Growth rules for irregular architected materials with programmable properties},
    volume = {377},
    url = {https://www.science.org/doi/full/10.1126/science.abn1459},
    doi = {10.1126/science.abn1459},
    urldate = {2025-12-17},
    journal = {Science},
    author = {Liu, Ke and Sun, Rachel and Daraio, Chiara},
    year = {2022},
    pages = {975--981},
}

@article{zaiser_disordered_2023,
    title = {Disordered mechanical metamaterials},
    volume = {5},
    copyright = {Copyright Nature Publishing Group Nov 2023},
    url = {https://www.proquest.com/docview/3226305849/abstract/17E8BE10CACE441BPQ/1},
    doi = {10.1038/s42254-023-00639-3},
    number = {11},
    urldate = {2025-12-17},
    journal = {Nature Reviews. Physics},
    author = {Zaiser, Michael and Zapperi, Stefano},
    year = {2023},
    pages = {679--688},
}

@article{shaikeea_toughness_2022,
    title = {The toughness of mechanical metamaterials},
    volume = {21},
    copyright = {2022 The Author(s), under exclusive licence to Springer Nature Limited},
    issn = {1476-4660},
    url = {https://www.nature.com/articles/s41563-021-01182-1},
    doi = {10.1038/s41563-021-01182-1},
    number = {3},
    urldate = {2025-12-17},
    journal = {Nature Materials},
    author = {Shaikeea, Angkur Jyoti Dipanka and Cui, Huachen and O’Masta, Mark and Zheng, Xiaoyu Rayne and Deshpande, Vikram Sudhir},
    year = {2022},
    pages = {297--304},
}

@article{coulais_combinatorial_2016,
    title = {Combinatorial design of textured mechanical metamaterials},
    volume = {535},
    copyright = {2016 Springer Nature Limited},
    issn = {1476-4687},
    url = {https://www.nature.com/articles/nature18960},
    doi = {10.1038/nature18960},
    number = {7613},
    urldate = {2025-12-17},
    journal = {Nature},
    author = {Coulais, Corentin and Teomy, Eial and de Reus, Koen and Shokef, Yair and van Hecke, Martin},
    year = {2016},
    pages = {529--532},
}

@article{walker_computational_2024,
    title = {Computational design of {4D} printed shape morphing lattices undergoing large deformation},
    volume = {33},
    issn = {0964-1726},
    url = {https://dx.doi.org/10.1088/1361-665X/ad8a31},
    doi = {10.1088/1361-665X/ad8a31},
    number = {11},
    urldate = {2025-01-17},
    journal = {Smart Materials and Structures},
    author = {Walker, Andreas and Shea, Kristina},
    year = {2024},
    pages = {115047},
}

@article{bonfanti_automatic_2020,
    title = {Automatic design of mechanical metamaterial actuators},
    volume = {11},
    copyright = {2020 The Author(s)},
    issn = {2041-1723},
    url = {https://www.nature.com/articles/s41467-020-17947-2},
    doi = {10.1038/s41467-020-17947-2},
    language = {en},
    number = {1},
    urldate = {2025-12-17},
    journal = {Nature Communications},
    author = {Bonfanti, Silvia and Guerra, Roberto and Font-Clos, Francesc and Rayneau-Kirkhope, Daniel and Zapperi, Stefano},
    year = {2020},
    pages = {4162},
}

@article{alnaes_unified_2014,
    title = {Unified form language: {A} domain-specific language for weak formulations of partial differential equations},
    volume = {40},
    issn = {0098-3500},
    shorttitle = {Unified form language},
    url = {https://dl.acm.org/doi/10.1145/2566630},
    doi = {10.1145/2566630},
    number = {2},
    urldate = {2026-02-16},
    journal = {ACM Trans. Math. Softw.},
    author = {Alnæs, Martin S. and Logg, Anders and Ølgaard, Kristian B. and Rognes, Marie E. and Wells, Garth N.},
    year = {2014},
    pages = {9:1--9:37},
}

@article{scroggs_construction_2022,
    title = {Construction of {Arbitrary} {Order} {Finite} {Element} {Degree}-of-{Freedom} {Maps} on {Polygonal} and {Polyhedral} {Cell} {Meshes}},
    volume = {48},
    issn = {0098-3500},
    url = {https://dl.acm.org/doi/10.1145/3524456},
    doi = {10.1145/3524456},
    number = {2},
    urldate = {2026-02-16},
    journal = {ACM Trans. Math. Softw.},
    author = {Scroggs, Matthew W. and Dokken, Jørgen S. and Richardson, Chris N. and Wells, Garth N.},
    year = {2022},
    pages = {18:1--18:23},
}

@misc{baratta_dolfinx_2023,
    title = {{DOLFINx}: {The} next generation {FEniCS} problem solving environment},
    shorttitle = {{DOLFINx}},
    url = {https://zenodo.org/records/10447666},
    doi = {10.5281/zenodo.10447666},
    urldate = {2026-02-16},
    publisher = {Zenodo},
    author = {Baratta, Igor A. and Dean, Joseph P. and Dokken, Jørgen S. and Habera, Michal and Hale, Jack S. and Richardson, Chris N. and Rognes, Marie E. and Scroggs, Matthew W. and Sime, Nathan and Wells, Garth N.},
    year = {2023},
}

@Misc{            petsc-web-page,
  author        = {Satish Balay and Shrirang Abhyankar and Mark~F. Adams and Steven Benson and Jed
                  Brown and Peter Brune and Kris Buschelman and Emil~M. Constantinescu and Lisandro
                  Dalcin and Alp Dener and Victor Eijkhout and Jacob Faibussowitsch and William~D.
                  Gropp and V\'{a}clav Hapla and Tobin Isaac and Pierre Jolivet and Dmitry Karpeev
                  and Dinesh Kaushik and Matthew~G. Knepley and Fande Kong and Scott Kruger and
                  Dave~A. May and Lois Curfman McInnes and Richard Tran Mills and Lawrence Mitchell
                  and Todd Munson and Jose~E. Roman and Karl Rupp and Patrick Sanan and Jason Sarich
                  and Barry~F. Smith and Stefano Zampini and Hong Zhang and Hong Zhang and Junchao
                  Zhang},
  title         = {{PETS}c {W}eb page},
  url           = {https://petsc.org/},
  howpublished  = {\url{https://petsc.org/}},
  year          = {2025}
}

@TechReport{      petsc-user-ref,
  author        = {Satish Balay and Shrirang Abhyankar and Mark~F. Adams and Steven Benson and Jed
                  Brown and Peter Brune and Kris Buschelman and Emil Constantinescu and Lisandro
                  Dalcin and Alp Dener and Victor Eijkhout and Jacob Faibussowitsch and William~D.
                  Gropp and V\'{a}clav Hapla and Tobin Isaac and Pierre Jolivet and Dmitry Karpeev
                  and Dinesh Kaushik and Matthew~G. Knepley and Fande Kong and Scott Kruger and
                  Dave~A. May and Lois Curfman McInnes and Richard Tran Mills and Lawrence Mitchell
                  and Todd Munson and Jose~E. Roman and Karl Rupp and Patrick Sanan and Jason Sarich
                  and Barry~F. Smith and Hansol Suh and Stefano Zampini and Hong Zhang and Hong Zhang
                  and Junchao Zhang},
  title         = {{PETSc/TAO} Users Manual},
  institution   = {Argonne National Laboratory},
  number        = {ANL-21/39 - Revision 3.24},
  doi           = {10.2172/2998643},
  year          = {2025}
}

@InProceedings{   petsc-efficient,
  author        = {Satish Balay and William~D. Gropp and Lois Curfman McInnes and Barry~F. Smith},
  title         = {Efficient Management of Parallelism in Object Oriented Numerical Software
                  Libraries},
  booktitle     = {Modern Software Tools in Scientific Computing},
  editor        = {E. Arge and A.~M. Bruaset and H.~P. Langtangen},
  publisher     = {Birkh{\"{a}}user Press},
  pages         = {163--202},
  year          = {1997}
}

@Article{         dalcinpazklercosimo2011,
  title         = {Parallel distributed computing using Python},
  author        = {Lisandro D. Dalcin and Rodrigo R. Paz and Pablo A. Kler and Alejandro Cosimo},
  journal       = {Advances in Water Resources},
  volume        = {34},
  number        = {9},
  pages         = {1124 - 1139},
  issn          = {0309-1708},
  doi           = {10.1016/j.advwatres.2011.04.013},
  year          = {2011}
}

@article{geuzaine_gmsh_2009,
    title = {Gmsh: {A} 3-{D} finite element mesh generator with built-in pre- and post-processing facilities},
    volume = {79},
    copyright = {Copyright © 2009 John Wiley \& Sons, Ltd.},
    issn = {1097-0207},
    shorttitle = {Gmsh},
    url = {https://onlinelibrary.wiley.com/doi/abs/10.1002/nme.2579},
    doi = {10.1002/nme.2579},
    number = {11},
    urldate = {2026-02-16},
    journal = {International Journal for Numerical Methods in Engineering},
    author = {Geuzaine, Christophe and Remacle, Jean-François},
    year = {2009},
    pages = {1309--1331},
}

@article{bastek_inverse_2023,
    title = {Inverse design of nonlinear mechanical metamaterials via video denoising diffusion models},
    volume = {5},
    copyright = {2023 The Author(s)},
    issn = {2522-5839},
    url = {https://www.nature.com/articles/s42256-023-00762-x},
    doi = {10.1038/s42256-023-00762-x},
    number = {12},
    urldate = {2024-09-08},
    journal = {Nature Machine Intelligence},
    publisher = {Nature Publishing Group},
    author = {Bastek, Jan-Hendrik and Kochmann, Dennis M.},
    year = {2023},
    pages = {1466--1475},
}

@misc{ronneberger_u-net_2015,
    title = {U-{Net}: {Convolutional} {Networks} for {Biomedical} {Image} {Segmentation}},
    shorttitle = {U-{Net}},
    url = {http://arxiv.org/abs/1505.04597},
    doi = {10.48550/arXiv.1505.04597},
    urldate = {2026-02-18},
    publisher = {arXiv},
    author = {Ronneberger, Olaf and Fischer, Philipp and Brox, Thomas},
    year = {2015},
    note = {arXiv:1505.04597 [cs]},
}

@misc{paszke_pytorch_2019,
    title = {{PyTorch}: {An} {Imperative} {Style}, {High}-{Performance} {Deep} {Learning} {Library}},
    shorttitle = {{PyTorch}},
    url = {http://arxiv.org/abs/1912.01703},
    doi = {10.48550/arXiv.1912.01703},
    urldate = {2026-02-18},
    publisher = {arXiv},
    author = {Paszke, Adam and Gross, Sam and Massa, Francisco and Lerer, Adam and Bradbury, James and Chanan, Gregory and Killeen, Trevor and Lin, Zeming and Gimelshein, Natalia and Antiga, Luca and Desmaison, Alban and Köpf, Andreas and Yang, Edward and DeVito, Zach and Raison, Martin and Tejani, Alykhan and Chilamkurthy, Sasank and Steiner, Benoit and Fang, Lu and Bai, Junjie and Chintala, Soumith},
    year = {2019},
}

@misc{ho_denoising_2020,
    title = {Denoising {Diffusion} {Probabilistic} {Models}},
    url = {http://arxiv.org/abs/2006.11239},
    doi = {10.48550/arXiv.2006.11239},
    urldate = {2025-03-01},
    publisher = {arXiv},
    author = {Ho, Jonathan and Jain, Ajay and Abbeel, Pieter},
    year = {2020},
}

@misc{rombach_high-resolution_2022,
    title = {High-{Resolution} {Image} {Synthesis} with {Latent} {Diffusion} {Models}},
    url = {http://arxiv.org/abs/2112.10752},
    doi = {10.48550/arXiv.2112.10752},
    urldate = {2025-04-03},
    publisher = {arXiv},
    author = {Rombach, Robin and Blattmann, Andreas and Lorenz, Dominik and Esser, Patrick and Ommer, Björn},
    year = {2022},
}

@article{kollmann_deep_2020,
    title = {Deep learning for topology optimization of {2D} metamaterials},
    volume = {196},
    issn = {0264-1275},
    url = {https://www.sciencedirect.com/science/article/pii/S026412752030633X},
    doi = {10.1016/j.matdes.2020.109098},
    urldate = {2026-03-10},
    journal = {Materials \& Design},
    author = {Kollmann, Hunter T. and Abueidda, Diab W. and Koric, Seid and Guleryuz, Erman and Sobh, Nahil A.},
    year = {2020},
    pages = {109098},
}

@article{mao_designing_2020,
    title = {Designing complex architectured materials with generative adversarial networks},
    volume = {6},
    url = {https://www.science.org/doi/10.1126/sciadv.aaz4169},
    doi = {10.1126/sciadv.aaz4169},
    number = {17},
    urldate = {2026-03-10},
    journal = {Science Advances},
    publisher = {American Association for the Advancement of Science},
    author = {Mao, Yunwei and He, Qi and Zhao, Xuanhe},
    year = {2020},
    pages = {eaaz4169},
}

@inproceedings{sorkine_laplacian_2004,
    address = {New York, NY, USA},
    series = {{SGP} '04},
    title = {Laplacian surface editing},
    isbn = {978-3-905673-13-5},
    url = {https://dl.acm.org/doi/10.1145/1057432.1057456},
    doi = {10.1145/1057432.1057456},
    urldate = {2025-05-06},
    booktitle = {Proceedings of the 2004 {Eurographics}/{ACM} {SIGGRAPH} symposium on {Geometry} processing},
    publisher = {Association for Computing Machinery},
    author = {Sorkine, O. and Cohen-Or, D. and Lipman, Y. and Alexa, M. and Rössl, C. and Seidel, H.-P.},
    year = {2004},
    pages = {175--184},
}

@article{igarashi_as-rigid-as-possible_2005,
    title = {As-rigid-as-possible shape manipulation},
    volume = {24},
    issn = {0730-0301, 1557-7368},
    url = {https://dl.acm.org/doi/10.1145/1073204.1073323},
    doi = {10.1145/1073204.1073323},
    language = {en},
    number = {3},
    urldate = {2025-06-13},
    journal = {ACM Transactions on Graphics},
    author = {Igarashi, Takeo and Moscovich, Tomer and Hughes, John F.},
    year = {2005},
    pages = {1134--1141},
}

@article{sim_electromagnetic_nodate,
    title = {Electromagnetic ({EM})‐{Driven} {Functional} {Materials}},
    url = {https://advanced.onlinelibrary.wiley.com/doi/10.1002/adma.202521268},
    doi = {10.1002/adma.202521268},
    journal = {Advanced Materials},
    year = {2026},
    volume = {38},
    number = {11},
    pages = {e21268},
    urldate = {2026-03-24},
    author = {Sim, Jay and Lu, Lu and Zhao, Ruike Renee},
}

@misc{ho_imagen_2022,
    title = {Imagen {Video}: {High} {Definition} {Video} {Generation} with {Diffusion} {Models}},
    shorttitle = {Imagen {Video}},
    url = {http://arxiv.org/abs/2210.02303},
    doi = {10.48550/arXiv.2210.02303},
    urldate = {2026-03-31},
    publisher = {arXiv},
    author = {Ho, Jonathan and Chan, William and Saharia, Chitwan and Whang, Jay and Gao, Ruiqi and Gritsenko, Alexey and Kingma, Diederik P. and Poole, Ben and Norouzi, Mohammad and Fleet, David J. and Salimans, Tim},
    year = {2022},
}

@misc{selig_uiuc_1996,
    title = {{UIUC} {Airfoil} {Data} {Site}},
    url = {http://www.worldcat.org/oclc/44283102},
    urldate = {2026-03-31},
    publisher = {University of Illinois Urbana-Champaign},
    author = {Selig, Michael S.},
    year = {1996},
}

@article{chen_airfoil_2020,
    title = {Airfoil {Design} {Parameterization} and {Optimization} {Using} {Bézier} {Generative} {Adversarial} {Networks}},
    volume = {58},
    issn = {0001-1452, 1533-385X},
    url = {https://arc.aiaa.org/doi/10.2514/1.J059317},
    doi = {10.2514/1.J059317},
    number = {11},
    urldate = {2026-03-31},
    journal = {AIAA Journal},
    author = {Chen, Wei and Chiu, Kevin and Fuge, Mark D.},
    year = {2020},
    pages = {4723--4735},
}

@article{dudek_shape-morphing_2025,
    title = {Shape-morphing metamaterials},
    volume = {10},
    copyright = {2025 Springer Nature Limited},
    issn = {2058-8437},
    url = {https://www.nature.com/articles/s41578-025-00828-9},
    doi = {10.1038/s41578-025-00828-9},
    language = {en},
    number = {10},
    urldate = {2026-04-11},
    journal = {Nature Reviews Materials},
    publisher = {Nature Publishing Group},
    author = {Dudek, Krzysztof K. and Kadic, Muamer and Coulais, Corentin and Bertoldi, Katia},
    year = {2025},
    pages = {783--798},
}

@article{xia_electrochemically_2019,
    title = {Electrochemically reconfigurable architected materials},
    volume = {573},
    copyright = {2019 The Author(s), under exclusive licence to Springer Nature Limited},
    issn = {1476-4687},
    url = {https://www.nature.com/articles/s41586-019-1538-z},
    doi = {10.1038/s41586-019-1538-z},
    language = {en},
    number = {7773},
    urldate = {2026-04-11},
    journal = {Nature},
    publisher = {Nature Publishing Group},
    author = {Xia, Xiaoxing and Afshar, Arman and Yang, Heng and Portela, Carlos M. and Kochmann, Dennis M. and Di Leo, Claudio V. and Greer, Julia R.},
    month = sep,
    year = {2019},
    pages = {205--213},
}

@article{alapan_reprogrammable_2020,
    title = {Reprogrammable shape morphing of magnetic soft machines},
    volume = {6},
    url = {https://www.science.org/doi/full/10.1126/sciadv.abc6414},
    doi = {10.1126/sciadv.abc6414},
    number = {38},
    urldate = {2026-04-11},
    journal = {Science Advances},
    publisher = {American Association for the Advancement of Science},
    author = {Alapan, Yunus and Karacakol, Alp C. and Guzelhan, Seyda N. and Isik, Irem and Sitti, Metin},
    month = sep,
    year = {2020},
    pages = {eabc6414},
}

@article{kim_ferromagnetic_2019,
    title = {Ferromagnetic soft continuum robots},
    volume = {4},
    issn = {2470-9476},
    url = {https://www.science.org/doi/10.1126/scirobotics.aax7329},
    doi = {10.1126/scirobotics.aax7329},
    language = {en},
    number = {33},
    urldate = {2026-04-11},
    journal = {Science Robotics},
    author = {Kim, Yoonho and Parada, German A. and Liu, Shengduo and Zhao, Xuanhe},
    month = aug,
    year = {2019},
    pages = {eaax7329},
}

@article{wang_performance_2024,
    title = {Performance metrics for shape-morphing devices},
    volume = {9},
    copyright = {2024 Springer Nature Limited},
    issn = {2058-8437},
    url = {https://www.nature.com/articles/s41578-024-00714-w},
    doi = {10.1038/s41578-024-00714-w},
    language = {en},
    number = {10},
    urldate = {2026-04-11},
    journal = {Nature Reviews Materials},
    publisher = {Nature Publishing Group},
    author = {Wang, Jue and Chortos, Alex},
    month = oct,
    year = {2024},
    pages = {738--751},
}

@article{bai_dynamically_2022,
    title = {A dynamically reprogrammable surface with self-evolving shape morphing},
    volume = {609},
    copyright = {2022 The Author(s), under exclusive licence to Springer Nature Limited},
    issn = {1476-4687},
    url = {https://www.nature.com/articles/s41586-022-05061-w},
    doi = {10.1038/s41586-022-05061-w},
    language = {en},
    number = {7928},
    urldate = {2026-04-12},
    journal = {Nature},
    publisher = {Nature Publishing Group},
    author = {Bai, Yun and Wang, Heling and Xue, Yeguang and Pan, Yuxin and Kim, Jin-Tae and Ni, Xinchen and Liu, Tzu-Li and Yang, Yiyuan and Han, Mengdi and Huang, Yonggang and Rogers, John A. and Ni, Xiaoyue},
    month = sep,
    year = {2022},
    pages = {701--708},
}

@article{meeussen_multistable_2023,
    title = {Multistable sheets with rewritable patterns for switchable shape-morphing},
    volume = {621},
    copyright = {2023 The Author(s), under exclusive licence to Springer Nature Limited},
    issn = {1476-4687},
    url = {https://www.nature.com/articles/s41586-023-06353-5},
    doi = {10.1038/s41586-023-06353-5},
    language = {en},
    number = {7979},
    urldate = {2026-04-12},
    journal = {Nature},
    publisher = {Nature Publishing Group},
    author = {Meeussen, A. S. and van Hecke, M.},
    month = sep,
    year = {2023},
    pages = {516--520},
}

@article{siefert_bio-inspired_2019,
    title = {Bio-inspired pneumatic shape-morphing elastomers},
    volume = {18},
    copyright = {2018 The Author(s), under exclusive licence to Springer Nature Limited},
    issn = {1476-4660},
    url = {https://www.nature.com/articles/s41563-018-0219-x},
    doi = {10.1038/s41563-018-0219-x},
    language = {en},
    number = {1},
    urldate = {2026-04-12},
    journal = {Nature Materials},
    publisher = {Nature Publishing Group},
    author = {Siéfert, Emmanuel and Reyssat, Etienne and Bico, José and Roman, Benoît},
    month = jan,
    year = {2019},
    pages = {24--28},
}

@article{lu_making_2014,
    title = {Making strong nanomaterials ductile with gradients},
    volume = {345},
    url = {https://www.science.org/doi/full/10.1126/science.1255940},
    doi = {10.1126/science.1255940},
    number = {6203},
    urldate = {2026-04-12},
    journal = {Science},
    publisher = {American Association for the Advancement of Science},
    author = {Lu, K.},
    month = sep,
    year = {2014},
    pages = {1455--1456},
}

@article{xu_planar_2016,
    title = {Planar gradient metamaterials},
    volume = {1},
    copyright = {2016 Macmillan Publishers Limited},
    issn = {2058-8437},
    url = {https://www.nature.com/articles/natrevmats201667},
    doi = {10.1038/natrevmats.2016.67},
    language = {en},
    number = {12},
    urldate = {2026-04-12},
    journal = {Nature Reviews Materials},
    publisher = {Nature Publishing Group},
    author = {Xu, Yadong and Fu, Yangyang and Chen, Huanyang},
    month = oct,
    year = {2016},
    pages = {16067},
}

@article{ma_deep_2022,
    title = {Deep {Learning}-{Accelerated} {Designs} of {Tunable} {Magneto}-{Mechanical} {Metamaterials}},
    volume = {14},
    copyright = {https://doi.org/10.15223/policy-029},
    issn = {1944-8244, 1944-8252},
    url = {https://pubs.acs.org/doi/10.1021/acsami.2c09052},
    doi = {10.1021/acsami.2c09052},
    language = {en},
    number = {29},
    urldate = {2026-04-16},
    journal = {ACS Applied Materials \& Interfaces},
    author = {Ma, Chunping and Chang, Yilong and Wu, Shuai and Zhao, Ruike Renee},
    month = jul,
    year = {2022},
    pages = {33892--33902},
}

@article{liwei_2025,
author = {Liwei Wang  and Alexander L. Evenchik  and Jared M. Yang  and Ryan L. Truby  and Wei Chen },
title = {Autonomous codesign and fabrication of multistimuli-responsive material systems},
journal = {Science Advances},
volume = {11},
number = {37},
pages = {eadx4409},
year = {2025},
doi = {10.1126/sciadv.adx4409},
URL = {https://www.science.org/doi/abs/10.1126/sciadv.adx4409},
}

@inproceedings{farmaga_evaluation_2011,
    title = {Evaluation of computational complexity of finite element analysis},
    url = {https://ieeexplore.ieee.org/document/5744437/},
    urldate = {2026-01-27},
    booktitle = {2011 11th {International} {Conference} {The} {Experience} of {Designing} and {Application} of {CAD} {Systems} in {Microelectronics} ({CADSM})},
    author = {Farmaga, Ihor and Shmigelskyi, Petro and Spiewak, Piotr and Ciupinski, Lukasz},
    month = feb,
    year = {2011},
    pages = {213--214},
}

@article{mm_mesh_2016,
    title = {Mesh {Deformation} {Approaches} – {A} {Survey}},
    volume = {7},
    issn = {20900902},
    url = {https://www.omicsonline.com/open-access/mesh-deformation-approaches--a-survey-2090-0902-1000181.php?aid=76056},
    doi = {10.4172/2090-0902.1000181},
    language = {en},
    number = {2},
    urldate = {2026-03-02},
    journal = {Journal of Physical Mathematics},
    author = {Selim, M. M. and Koomullil, R. P.},
    year = {2016},
}

@article{zhang_temperature-responsive_2024,
    title = {Temperature-responsive metamaterials made of highly sensitive thermostat metal strips},
    volume = {10},
    url = {https://www.science.org/doi/10.1126/sciadv.ads0892},
    doi = {10.1126/sciadv.ads0892},
    number = {49},
    urldate = {2026-04-03},
    journal = {Science Advances},
    publisher = {American Association for the Advancement of Science},
    author = {Zhang, Yi and Jiang, Wei Zhong and Pan, Yang and Teng, Xing Chi and Xu, Hang Hang and Yan, Han and Ni, Xi Hai and Dong, Jun and Han, Dong and Chen, Wei Qiu and Yang, Jie and Xie, Yi Min and Lu, Yang and Ren, Xin},
    year = {2024},
    pages = {eads0892},
}

@article{zhang_shape_2022,
    title = {Shape morphing of plastic films},
    volume = {13},
    copyright = {2022 The Author(s)},
    issn = {2041-1723},
    url = {https://www.nature.com/articles/s41467-022-34844-y},
    doi = {10.1038/s41467-022-34844-y},
    number = {1},
    urldate = {2026-04-03},
    journal = {Nature Communications},
    publisher = {Nature Publishing Group},
    author = {Zhang, Feilong and Li, Dong and Wang, Changxian and Liu, Zhihua and Yang, Man and Cui, Zequn and Yi, Junqi and Wang, Ming and Jiang, Ying and Lv, Zhisheng and Wang, Shutao and Gao, Huajian and Chen, Xiaodong},
    year = {2022},
    pages = {7294},
}

@misc{vaswani_attention_2023,
    title = {Attention {Is} {All} {You} {Need}},
    url = {http://arxiv.org/abs/1706.03762},
    doi = {10.48550/arXiv.1706.03762},
    urldate = {2026-04-06},
    publisher = {arXiv},
    author = {Vaswani, Ashish and Shazeer, Noam and Parmar, Niki and Uszkoreit, Jakob and Jones, Llion and Gomez, Aidan N. and Kaiser, Lukasz and Polosukhin, Illia},
    year = {2023},
}

@misc{he_deep_2015,
    title = {Deep {Residual} {Learning} for {Image} {Recognition}},
    url = {http://arxiv.org/abs/1512.03385},
    doi = {10.48550/arXiv.1512.03385},
    urldate = {2026-04-06},
    publisher = {arXiv},
    author = {He, Kaiming and Zhang, Xiangyu and Ren, Shaoqing and Sun, Jian},
    year = {2015},
}

@article{braden_surveyors_1986,
    title = {The {Surveyor}'s {Area} {Formula}},
    volume = {17},
    issn = {0746-8342},
    url = {https://doi.org/10.1080/07468342.1986.11972974},
    doi = {10.1080/07468342.1986.11972974},
    number = {4},
    urldate = {2026-04-20},
    journal = {The College Mathematics Journal},
    publisher = {Taylor \& Francis},
    author = {Braden, Bart},
    month = sep,
    year = {1986},
    pages = {326--337},
}

@article{umeyama_least-squares_1991,
    title = {Least-squares estimation of transformation parameters between two point patterns},
    volume = {13},
    number = {4},
    url = {https://ieeexplore.ieee.org/document/88573},
    doi = {10.1109/34.88573},
    journal = {IEEE Transactions on Pattern Analysis and Machine Intelligence},
    author = {Umeyama, S.},
    year = {1991},
    pages = {376--380},
}

@article{schonemann_generalized_1966,
    title = {A generalized solution of the orthogonal procrustes problem},
    volume = {31},
    number = {1},
    url = {https://link.springer.com/article/10.1007/BF02289451},
    doi = {10.1007/BF02289451},
    journal = {Psychometrika},
    author = {Schönemann, P. H.},
    year = {1966},
    pages = {1--10},
}

@book{golub_matrix_2013,
    title = {Matrix {Computations}},
    edition = {4},
    author = {Golub, Gene H. and Van Loan, Charles F.},
    publisher = {Johns Hopkins University Press},
    address = {Baltimore},
    year = {2013},
    isbn = {978-1-4214-0794-4},
}

@article{george_nested_1973,
    title = {Nested dissection of a regular finite element mesh},
    author = {George, Alan},
    journal = {SIAM Journal on Numerical Analysis},
    volume = {10},
    number = {2},
    pages = {345--363},
    year = {1973},
    doi = {10.1137/0710032},
}

@article{brandt_multi-level_1977,
    title = {Multi-level adaptive solutions to boundary-value problems},
    author = {Brandt, Achi},
    journal = {Mathematics of Computation},
    volume = {31},
    number = {138},
    pages = {333--390},
    year = {1977},
    doi = {10.1090/S0025-5718-1977-0431719-X},
}

@article{amir_multigrid-cg_2014,
    title = {On multigrid-{CG} for efficient topology optimization},
    author = {Amir, Oded and Aage, Niels and Lazarov, Boyan S.},
    journal = {Structural and Multidisciplinary Optimization},
    volume = {49},
    number = {5},
    pages = {815--829},
    year = {2014},
    doi = {10.1007/s00158-013-1015-5},
}
\end{document}